\documentclass[pre,aps,superscriptaddress,nofootinbib,fleqn,twocolumn,showpacs]{revtex4}
\usepackage{amsmath,amssymb,graphicx}
\begin{document}

\title{Fractional derivatives of random walks: Time series with long-time memory}

\author{H.~E.~Roman}
\affiliation{Dipartimento~di~Fisica, Universit\`a~di~Milano-Bicocca,
             Piazza~della~Scienza~3, 20126~Milano, Italy}

\author{M.~Porto}
\affiliation{Institut~f\"ur~Festk\"orperphysik,
             Technische~Universit\"at~Darmstadt,
             Hochschulstr.~8, 64289~Darmstadt, Germany}

\date{\today}

\begin{abstract}
We review statistical properties of models generated by the application of a
(positive and negative order) fractional derivative operator to a standard random 
walk and show that the resulting stochastic walks display slowly-decaying 
autocorrelation functions. The relation between these correlated walks and the 
well-known fractionally integrated autoregressive (FIGARCH) models, commonly used 
in econometric studies, is discussed. The application of correlated random walks 
to simulate empirical financial times series is considered and compared with the 
predictions from FIGARCH and the simpler FIARCH processes. A comparison with empirical
data is performed.
\end{abstract}


\maketitle

%%%%%%%%%%%%%%%%%%%%%%%%%%%%%%%%%%%%%%%%%%%%%%%%%%%%%%%%%%%%%%%%%%%%%%%%%%%%%%%

\section{Introduction}

In recent years, there has been a surging interest in the application of
non-integer order differential operators to describe different types of temporal
and spatial anomalies displayed by complex systems
\cite{GionaRoman:1992,FracFokkerPlanck:1998,OldhamSpanier:1974,Podlubny:1999,%
Metzler:2000,Hilfer:2000,Gorenflo:2002,Gorenflo:2007,Scherer:2007}.
% FRACTIONAL CALCULUS
%\cite{GionaRoman:1992,Podlubny:1999,Metzler:2000,Hilfer:2000,Gorenflo:2002,Scherer:2007,Gorenflo:2007}
For example, a fractional diffusion equation suitably describes the asymptotic
behavior of independent random walkers on random fractal structures
\cite{GionaRoman:1992}, and also corresponding fractional Fokker-Planck
equations have been used to model anomalous behavior
\cite{FracFokkerPlanck:1998}. Several recent reviews can be found in literature
covering a vast field of applications (see e.g.
\cite{Metzler:2000,Hilfer:2000}). A treatment from a mathematical point of view
can be found in \cite{OldhamSpanier:1974,Podlubny:1999}.
Closely related to the present work is the subject of fractional derivative
operators as discussed within the framework of random walks
\cite{Gorenflo:2002,Gorenflo:2007} and anomalous transport phenomena
\cite{Scherer:2007}.

% ECONOPHYSICS
%\cite{onefactormodel,Gopikrishnan1998,Borland:1998,MantegnaStanley,Bouchaud:2001,%
%Podobnik2006,Roman2006,Scalas:2006,McCauley:2007}

Random walks play also an essential role in modeling the time evolution of stock prices. 
The literature in the field is huge and we just mention few papers related with the present 
work \cite{onefactormodel,Gopikrishnan1998,Borland:1998,MantegnaStanley,%
Bouchaud:2001,Podobnik2006,Roman2006,Scalas:2006,McCauley:2007}. The different features studied
include stock-index cross-correlations \cite{onefactormodel}, probability distribution functions
of log-returns \cite{Gopikrishnan1998,Borland:1998,MantegnaStanley,%
Bouchaud:2001,Podobnik2006,Roman2006,Scalas:2006,McCauley:2007}, leverage effect \cite{Bouchaud:2001},
stock-stock cross-correlations and market internal structure \cite{Roman2006}, tick-by-tick
behavior \cite{Scalas:2006}, non-stationarity issues \cite{McCauley:2007}.

% ARCH  and  % LONG-TIME MEMORY: FIGARCH
A succesful and widely used model to describe price variations is based on an autoregressive 
process with conditional heteroskedasticity (ARCH) due to Engle (see e.g.\ \cite{arch}), of which simple 
variants have been recently suggested \cite{RomanPorto2001,DosePortoRoman2003}. Further variants
of ARCH models have been recently discussed with regards to volatility \cite{tarch,tarch:Palatella2004}.
ARCH type models have been generalized to incorporate a long-time memory in the surrogate time
series, in an attempt to mimick the strong autocorrelations observed in volatility and absolute returns
\cite{Granger1980,Bollerslev:1986,Engle/Bollerslev:1986,Baillie1996,Zumbach2004,Tang2006}.

% HURST - FBM
%\cite{Hurst:1951,Mandelbrot:1968,fBm:math,Perez2006-fBm}
The issue of long-time autocorrelations, or long time memory, in time series is intimately related 
with the concept of Hurst exponent, introduced by Hurst many years ago to describe the 
persistence observed in the behavior of Nile floods \cite{Hurst:1951}. Motivated by these 
ideas, Mandelbrot introduced few years later a long-time memory model known as fractional Brownian 
motion (FBM), being stationary on all time scales, as an attempt to model such anomalous phenomena 
\cite{Mandelbrot:1968}. Rigorous mathematical aspects of FBM can be found in textbooks 
\cite{fBm:math}, the latter including also a discussion of L\'evy processes with long-time memory.
A recent work considers other aspects of the FBM model \cite{Perez2006-fBm}.

% FA
%\cite{Peng:1994,Kantelhardt:1995,PRL1998}
The problem of accurately determining the Hurst exponent of a time series has been the subject of
intense activity, started with the description of persistence in DNA sequences in which the detrended 
fluctuation analysis (DFA) has been introduced \cite{Peng:1994}. Further developments were achieved
based on Haar wavelets \cite{Kantelhardt:1995} and its generalization to higher orders \cite{PRL1998}. 
In the present work, we will make use of Haar wavelet techniques to analyse the surrogate time series 
and determining the associated Hurst exponents.

In this work, we apply a fractional derivation (and integration) operator to an uncorrelated random 
walk, obtained within a simple ARCH prescription, to generate a surrogate financial time series with 
uncorrelated variations on all time lags, but having a slowly decaying auto-correlation for the absolute 
variations, representing the absolute returns. The new process is denoted as fractional random walk 
ARCH (FRWARCH) and we show that it has finite second moments, in contrast to FIGARCH processes 
characterized by an infinite variance. We will conclude that FRWARCH can become a useful tool in 
econometrics applications.

The paper is organized as follows. In Sect.~\ref{append:fracderoper}, we briefly review the
fractional derivative operator and its main properties. In Sect.~\ref{append:fracrw}, we discuss 
the effects of applying a fractional derivative (and fractional integration) operator to a standard 
random walk. The scaling properties of the resulting fractional random walk are discussed. 
Sect.~\ref{frwarch} is devoted to FRWARCH, for which the associated probability distribution functions 
and auto-correlations are studied. In Sect.~\ref{figarch}, we review the widely used FIGARCH process 
and discuss several of its properties in comparison with those of FRWARCH. In Sect.~\ref{fiarch} we
briefly consider, for completeness, the simpler fractionally integrated ARCH models currently used 
in literature. In Sect.~\ref{empirical} we present a brief comparison of FRWARCH and FIGARCH with
empirical data to see how they actually perform in more realistic situations. Reference is made to
a simpler generalized ARCH model without long-range memory, but having a slow decaying autocorrelation 
function, to better understand the role of memory in the models presented here. Finally, Sect.~\ref{conclu} 
summarizes the main conclusions of the work.

%----------------------------------------------------------------------------------
\section{Fractional derivative operator}
\label{append:fracderoper}

Let us consider a time-dependent function $y(t)$ recorded at times $t=i\tau$, where
$0\le i\le N$ and $\tau$ is the time resolution. For simplicity, we will indicate the
associated time series as $y_i$. 

In the following, we study the fractional derivative (finite difference) operator of 
fractional order $\alpha$, $D_\tau^{(\alpha)}$, known in literature as the 
Gr\"unwald-Letnikov scheme (see e.g.\ \cite{Gorenflo:2002,Scherer:2007}), which is defined as
\begin{equation}
D_\tau^{(\alpha)} y_n={1\over \tau^\alpha} \sum_{i=0}^{n} (-1)^i {\alpha\choose i}
                       y_{n-i}
\label{eq:fracderoper}
\end{equation}
where 
\begin{equation}
{\alpha\choose i} = {\Gamma(\alpha+1) \over \Gamma(i+1) \Gamma(\alpha+1-i)}
\label{eq:binomcoeff}
\end{equation}
is the binomial coefficient and $\Gamma(x)$ is the Gamma function \cite{Abramowitz:1972}. 

For the numerical implementation of Eq.~(\ref{eq:fracderoper}), it is more convenient 
to work out an equivalent expression for the binomial coefficients in 
Eq.~(\ref{eq:binomcoeff}). This will also allow us to make contact with long-range memory 
models known from the financial literature. Using the properties of the $\Gamma(x)$ 
function \cite{Abramowitz:1972}, one can show that
\begin{equation}
{\alpha\choose i} = {\alpha(\alpha-1)(\alpha-2)...(\alpha-i+1)\over i!},
\label{eq:binomcoefffact}
\end{equation}
where the numerator is a polynomial of $i$th degree in $\alpha$. Using 
Eq.~(\ref{eq:binomcoefffact}), Eq.~(\ref{eq:fracderoper}) becomes
\begin{eqnarray}
D_\tau^{(\alpha)} y_n 
&=& {1\over \tau^\alpha} \left[~ y_n \right. - \alpha y_{n-1} + 
           {\alpha\over 2!} (\alpha-1) y_{n-2} - \dots \nonumber \\
&+& (-1)^n {\alpha\over n!} (\alpha-1)  \dots (\alpha-n+1) y_{0}  
           \left. \right].  
\label{eq:fracdernumer}
\end{eqnarray}

In what follows, we will consider values of $\alpha$ in the range $-1<\alpha<1$. 
Indeed, it is easy to see from Eq.~(\ref{eq:fracdernumer}) that the case $\alpha=1$ 
corresponds to the (Euler) first order derivative of $y_n$, i.e.,
\begin{equation}
D_\tau^{(1)} y_n = {1\over \tau} (y_n-y_{n-1}),
\label{eq:Euler1oper}
\end{equation}
while the value $\alpha=-1$ yields the integral of $y_n$,
\begin{equation}
D_\tau^{(-1)} y_n = \tau (y_n+y_{n-1}+y_{n-2}+\dots+y_0)=\sum_{i=0}^n y_i \tau.
\label{eq:Eulerm1oper}
\end{equation}

%%%%%%%%%%%%%%%%%%%%%%%%%%%%%%%%%%%%% FIGURE 1 %%%%%%%%%%%%%%%%%%%%%%%%%%%%%%%%%%%%%%%%
\begin{figure}[t]
\vspace{0.7cm}
\begin{center}
\includegraphics[width=6.5cm]{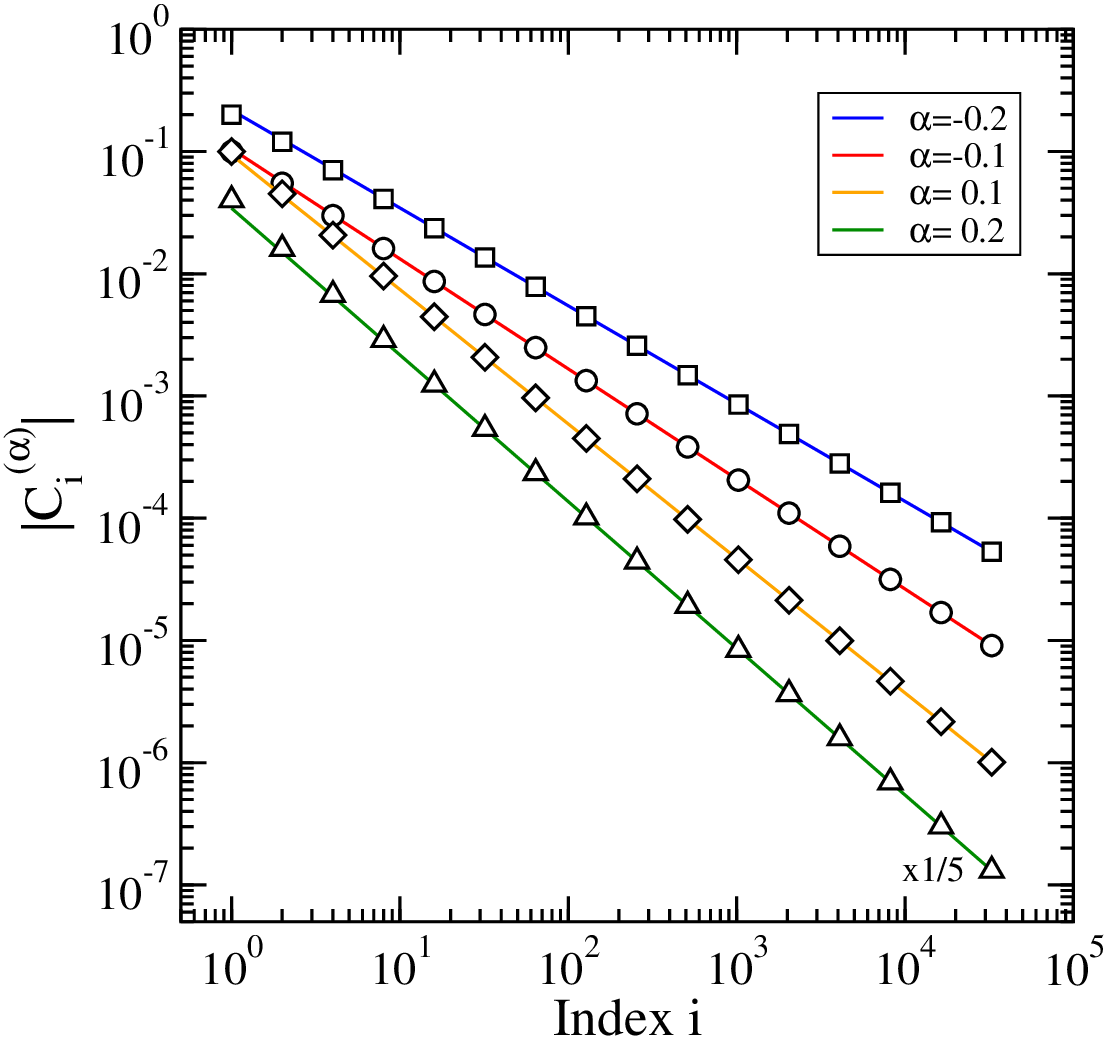}
\end{center}
\vspace{-0.5cm}
\caption{(color online) The absolute value of the coefficients $|C_i^{(\alpha)}|$ vs
         index $i$, for different values of the fractional order $\alpha=-0.2$ (squares),
	 -0.1 (circles), 0.1 (diamonds) and 0.2 (triangles). The symbols represent the 
	 numerically evaluated coefficients using Eq.~(\ref{eq:binomcoefffact}) and the 
       straight lines the asymptotic result, Eq.~(\ref{eq:asympcoeff}). The results 
       for $\alpha=0.2$ have been multiplied (down shifted) by a factor 1/5 for clarity.}
\label{fig:coeffnumer}
\end{figure}
%%%%%%%%%%%%%%%%%%%%%%%%%%%%%%%%%%%%%%%%%%%%%%%%%%%%%%%%%%%%%%%%%%%%%%%%%%%%%%%%%%%%%%%

\noindent
It is also instructive to derive the asymptotic behavior of the coefficients in the
expansion of the fractional operator. To do this, we write Eq.~(\ref{eq:binomcoefffact})
in the form
\begin{equation}
{\alpha\choose i} =  (-1)^i {(0-\alpha) (1-\alpha)(2-\alpha)...(i-1-\alpha)\over i!}.
\label{eq:binomcoefffact1}
\end{equation}
Now, using the relation $\prod_{j=0}^{i-1} (j+z)=\Gamma(i+z)/\Gamma(z)$ \cite{Abramowitz:1972} 
in Eq.~(\ref{eq:binomcoefffact1}), with $z=-\alpha$, the coefficient 
$C_i^{(\alpha)} \equiv (-1)^i {\alpha\choose i}$ can be written as
\begin{eqnarray}
C_i^{(\alpha)} &=& {\Gamma(i+z)\over \Gamma(z)~i!}        \nonumber \\
               &=& {\Gamma(i-\alpha)\over \Gamma(-\alpha) \Gamma(i+1)} = -
    {\alpha \over \Gamma(1-\alpha)} {\Gamma(i-\alpha)\over \Gamma(i+1)},
\label{eq:expancoeff}
\end{eqnarray}
where the known relations $\Gamma(1+z)=z\Gamma(z)$ and $i!=\Gamma(i+1)$ have been used.
Exploiting the large argument expansion $\Gamma(az+b)\simeq \exp(-az) (az)^{az+b-1/2}$,
one can obtain the asymptotic behavior of $C_i^{(\alpha)}$ for large $i$ as
\begin{equation}
C_i^{(\alpha)} \simeq - {\alpha\over \Gamma(1-\alpha)} i^{-(1+\alpha)}, \qquad i\gg1,
\label{eq:asympcoeff}
\end{equation}
which is positive for $\alpha<0$ and negative for $\alpha>0$. Results for $|C_i^{(\alpha)}|$ 
are plotted in Fig.~\ref{fig:coeffnumer} in double logarithmic scale for few representative 
values of $\alpha$. To be noted is the rather quick approach of the actual value of $C_i^{(\alpha)}$
to its asymptotic form Eq.~(\ref{eq:asympcoeff}). This feature can be used in numerical
calculations where $C_i^{(\alpha)}$ can be replaced by Eq.~(\ref{eq:asympcoeff}) when the 
difference between the two is sufficiently small.

\begin{figure}[t]
\vspace{0.7cm}
\begin{center}
\includegraphics[width=6.5cm]{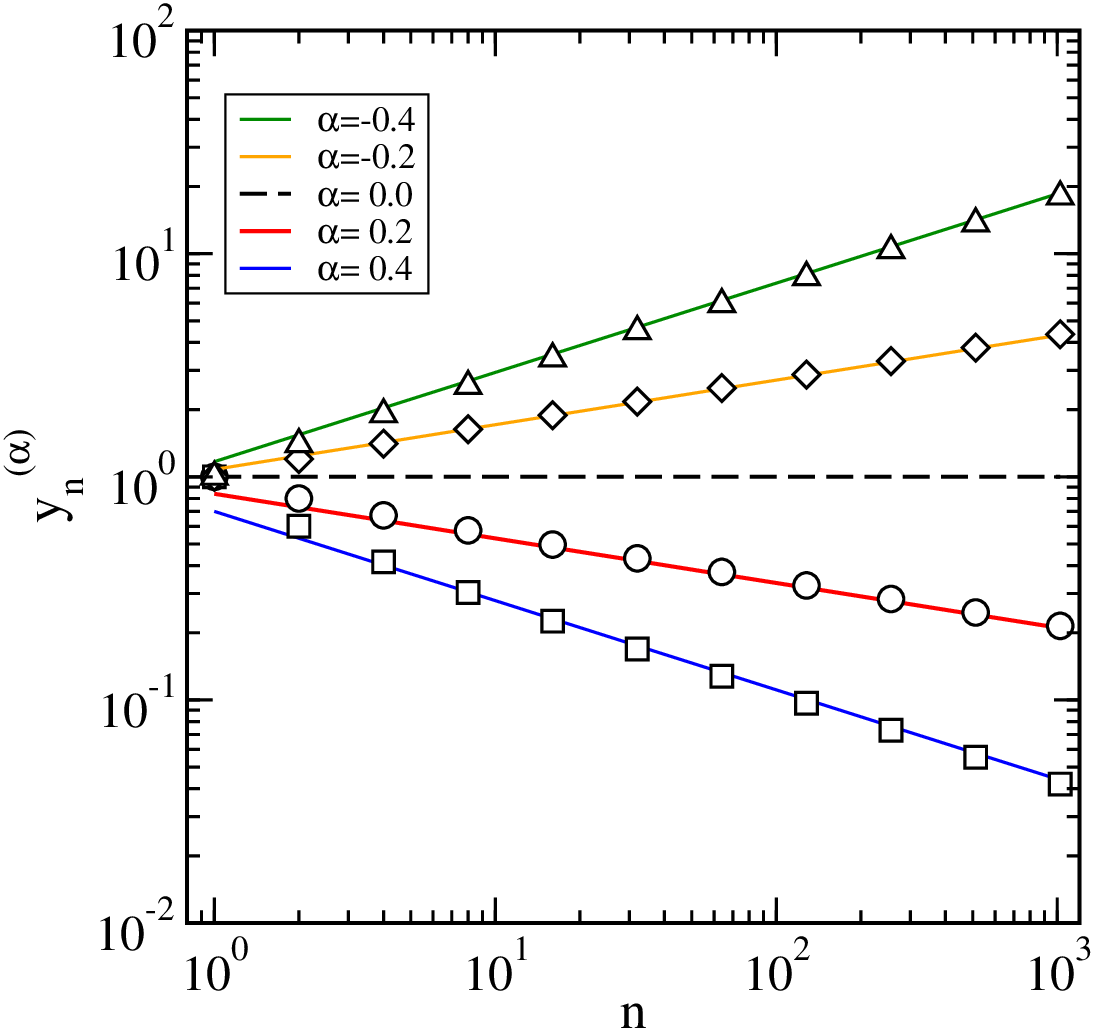}
\end{center}
\vspace{-0.5cm}
\caption{(color online) The fractional function $y_n^{(\alpha)}=D_\tau^{(\alpha)} y_n$ 
          vs index $n$, in the case $y_n=1$ (i.e.\ $\gamma=0$), $\tau=1$, for different 
	  values of the fractional order: $\alpha=0.4$ (squares), 0.2 (circles), -0.2 
	  (diamonds) and -0.4 (triangles). The symbols represent the numerical results 
	  using Eq.~(\ref{eq:fracdernumer}), and the straight lines the asymptotic form,
	  Eq.~(\ref{eq:exfracderoper}). The dashed line corresponds to the original
	  series $y_n$.}
\label{fig:examples}
\end{figure}

It is natural to refer to $D_\tau^{(\alpha)}$ as a fractional derivative operator for 
values $0<\alpha<1$ and as a fractional integral operator when $-1<\alpha<0$. This can 
be seen in the case that $y_n\simeq n^{\gamma}$. In fact, the action of the fractional 
operator, which will be denoted as
\begin{equation}
D_\tau^{(\alpha)} y_n \equiv y_n^{(\alpha)}={1\over \tau^{\alpha}} \sum_{i=0}^n C_i^{(\alpha)} y_{n-i},
\label{eq:deffracoper}
\end{equation}
can be estimated for large $n$ using the asymptotic behavior for $C_i^{(\alpha)}$, 
Eq.~(\ref{eq:asympcoeff}). Roughly, we can write 
$y_n^{(\alpha)}\simeq \int_1^n dx~x^{-(1+\alpha)}~(n-x)^\gamma$.
When $\gamma\ge 0$ and $n\gg1$, the largest contribution to the integral comes from the 
smallest values of $x$, yielding 
$y_n^{(\alpha)}\sim ~ n^{\gamma} \int_1^n dx~x^{-(1+\alpha)}\sim n^{\gamma-\alpha}$. Thus,
\begin{equation}
D_\tau^{(\alpha)} y_n \simeq n^{\gamma-\alpha}.
\label{eq:exfracderoper}
\end{equation}
We show in Fig.~\ref{fig:examples} numerical examples of the fractional operation
$D_1^{(\alpha)} y_n$ (i.e.\ $\tau=1$) in the case $y_n=1$ ($\gamma=0$), as an illustration
of the result Eq.~(\ref{eq:exfracderoper}). Note that for a constant function $y_n$,
the slope of the curves is just minus the order of the fractional operator. To be noted 
also is the rather quick approach of the fractional operator result 
Eq.~(\ref{eq:deffracoper}) to its asymptotic behavior Eq.~(\ref{eq:exfracderoper}),
already for small values of $n$.

%%%%%%%%%%%%%%%%%%%%%%%%%%%%%%%%%%%%% FIGURE 2 %%%%%%%%%%%%%%%%%%%%%%%%%%%%%%%%%%%%%%%%
%%%%%%%%%%%%%%%%%%%%%%%%%%%%%%%%%%%%%%%%%%%%%%%%%%%%%%%%%%%%%%%%%%%%%%%%%%%%%%%%%%%%%%%

%----------------------------------------------------------------------------------
\section{Fractional random walks}
\label{append:fracrw}

Let us consider the case in which $y_n$ represents a standard random walk (RW), and 
assume the time resolution $\tau=1$ for convenience. Specifically, $y_n$ is constructed 
as a sum of independent, equally distributed random numbers, $\eta$. Without loss of 
generality, let us assume the latter are drawn from a Gaussian distribution of zero mean 
and unit standard deviation, i.e.\ $\big<\eta\big>=0$ and $\big<\eta^2\big>=1$, yielding 
$\sigma_{\eta}=1$. Thus,
\begin{equation}
y_n= \sum_{i=1}^n \eta_i, \qquad 1\le n\le N,
\label{eq:gaussianrw}
\end{equation}
where we take $y_0=0$ for simplicity. We are interested in studying the behavior 
of the associated fractional random walks (FRW's), obtained by applying the fractional 
operator $D_1^{(\alpha)}$ to the random walk `profile' $y_n$, Eq.~(\ref{eq:gaussianrw}),
$D_1^{(\alpha)} y_n = y_n^{(\alpha)}$.

%%%%%%%%%%%%%%%%%%%%%%%%%%%%%%%%%%%%% FIGURE 3 %%%%%%%%%%%%%%%%%%%%%%%%%%%%%%%%%%%%%%%%
\begin{figure}[t]
\vspace{0.7cm}
\begin{center}
\includegraphics[width=6.5cm]{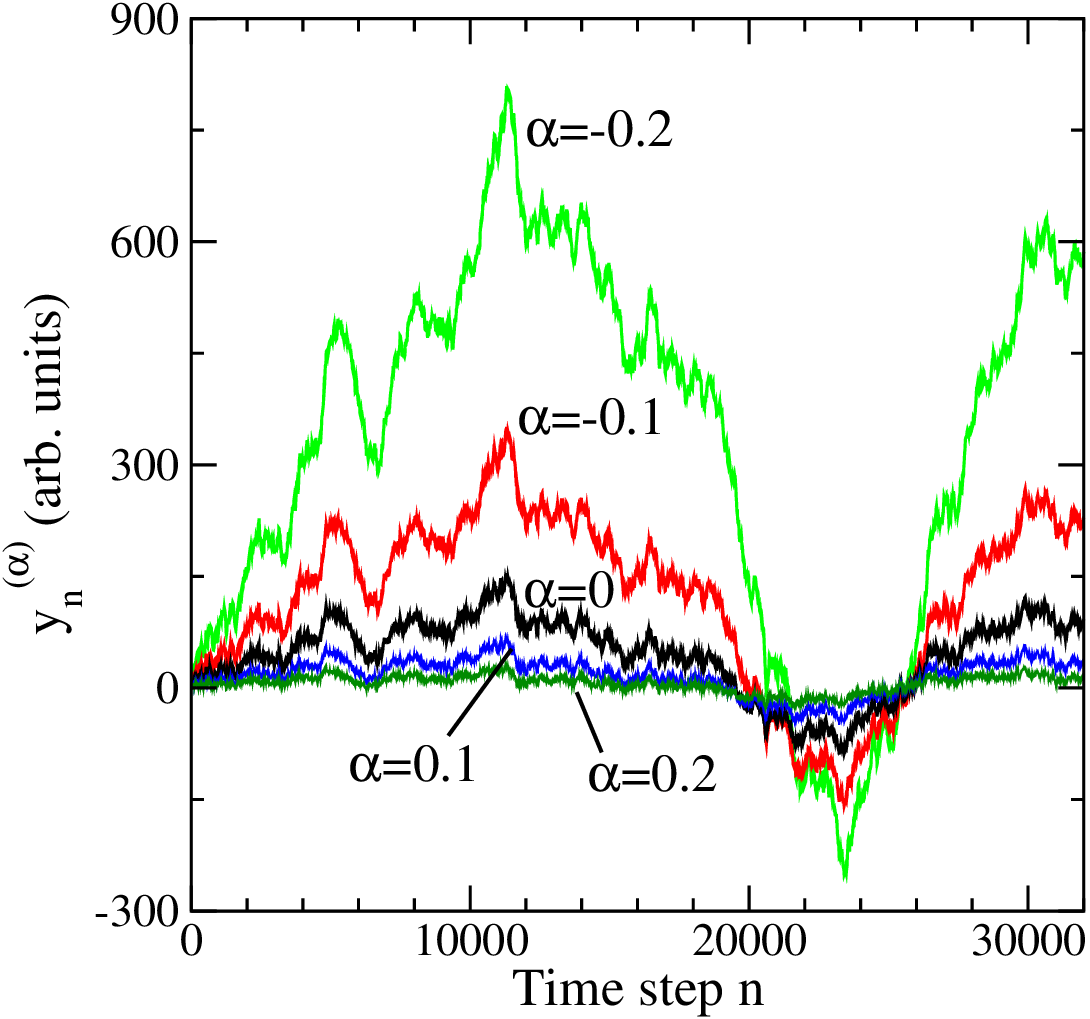}
\end{center}
\vspace{-0.5cm}
\caption{(color online) Fractional random walks: Plots of $y_n^{(\alpha)}$ versus time 
          step $n$, for selected values of $\alpha$ as indicated on the plot. The standard 
	  RW corresponding to $\alpha=0$, $y_n^{(0)}\equiv y_n$, is also shown for 
	  comparison. The walks consist of $N=32000$ steps each.}
\label{fig:fractrandwalks}
\end{figure}
%%%%%%%%%%%%%%%%%%%%%%%%%%%%%%%%%%%%%%%%%%%%%%%%%%%%%%%%%%%%%%%%%%%%%%%%%%%%%%%%%%%%%%%%

Examples of fractional walks $y_n^{(\alpha)}$ are illustrated in Fig.~\ref{fig:fractrandwalks}. 
As is apparent from the figure, the amplitude of the fractional walks increases for 
negative $\alpha$ with respect to the uncorrelated case $y_n$, while for positive $\alpha$ 
the amplitude of the walks gets smaller. The question is whether the fractional walks are 
also statistically different from the uncorrelated case in the sense that long-range 
autocorrelations are present. Before discussing this quest, let us consider next the 
distribution function of the increments of a FRW.

%----------------------------------------------------------------------------------
\subsection{Probability distribution functions}

The increments of a fractional random walk are denoted as 
\begin{equation}
\Delta y_n^{(\alpha)}=y_n^{(\alpha)}-y_{n-1}^{(\alpha)}.
\label{eq:defdeltaynalpha}
\end{equation}
For a standard random walk, as in Eq.~(\ref{eq:gaussianrw}), one simply has 
$\Delta y_n=\eta_n$, which is a local function of the time step $n$, i.e.\ independent 
of the previous values of $\eta_i$, $i<n$. In contrast, for any fractional order 
$\alpha$ ($\ne 0$), we find
\begin{eqnarray}
\Delta y_n^{(\alpha)} &=& \sum_{i=0}^{n} C_i^{(\alpha)} y_{n-i} - 
                          \sum_{i=0}^{n-1} C_i^{(\alpha)} y_{n-1-i}         \nonumber \\
                      &=& C_n^{(\alpha)} y_0 + \sum_{i=0}^{n-1} C_i^{(\alpha)} (y_{n-i}-y_{n-1-i})\nonumber \\
                      &=& \sum_{i=0}^{n-1} C_i^{(\alpha)} \eta_{n-i},
\label{eq:deltaynalpha}
\end{eqnarray}
where we have taken $y_0=0$. Thus, for fractional $\alpha$, the increments of the 
fractional random walk, $\Delta y_n^{(\alpha)}$, require the knowledge of the whole 
history of the walk, $\{\eta_i\}$, with $1\le i\le n$. This result suggests that FRW's 
may possess long-range autocorrelations, as we will indeed see below.

\begin{figure}[t]
\vspace{0.7cm}
\begin{center}
\includegraphics[width=6.5cm]{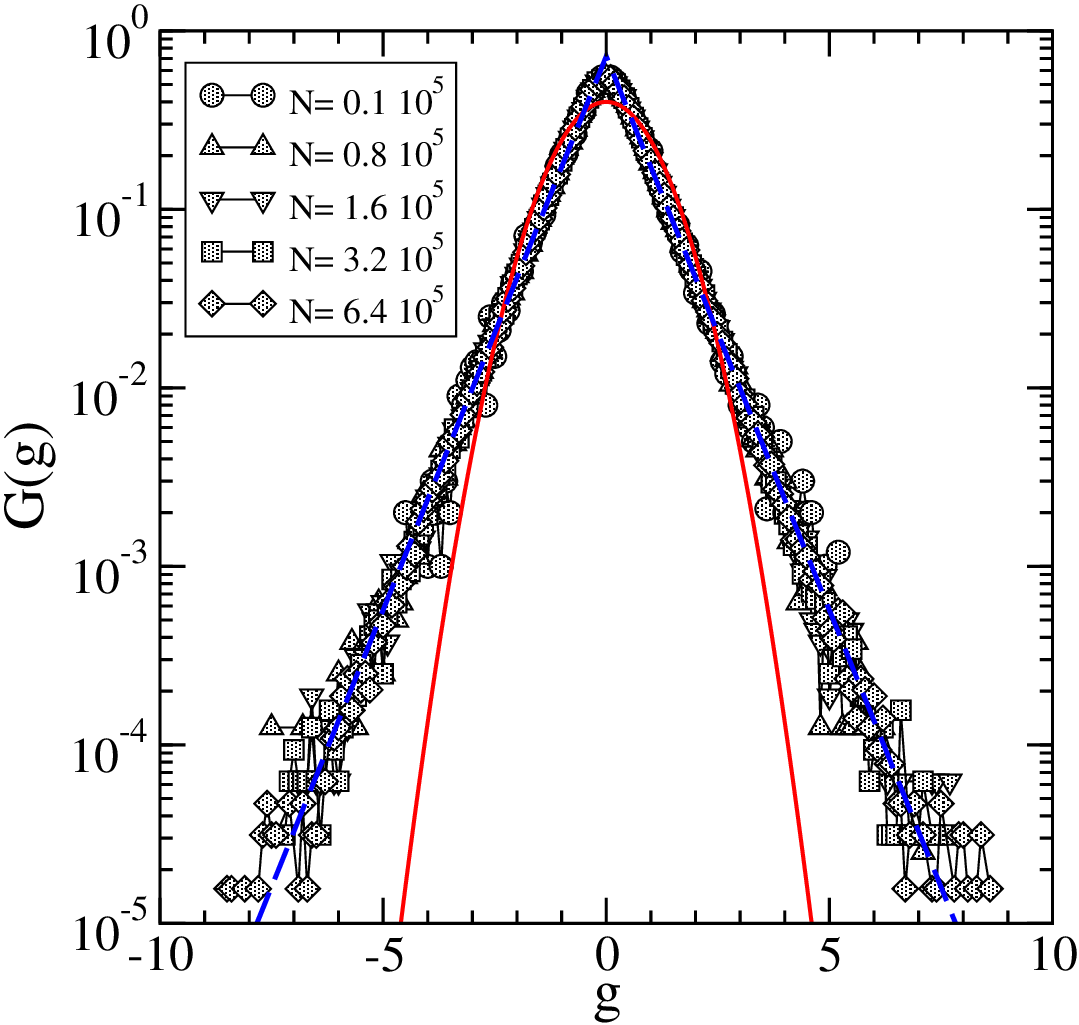}
\end{center}
\vspace{-0.5cm}
\caption{(color online) PDF, $G(g)$, for fractional random walks variations 
          $\Delta y_n^{(\alpha)}$ as a function of the scaled variable 
	  $g=(\Delta y_n^{(\alpha)}-\big<\Delta y_n^{(\alpha)}\big>)/\sigma_{\alpha}$
	  in the case $\alpha=-0.2$ (see Fig.~\ref{fig:fractrandwalks}), for series of
	  different length $N$ ($0\le n\le N$) indicated in the inset. The continuous 
	  line is the normal distribution, while the dashed line is the exponential
	  form $f(g)=A_1\exp(-|g|/A2)$, with $A_1=0.72$ and $A_2=0.7$.}
\label{fig:pdfwalks}
\end{figure}

The probability distribution function (PDF) of the increments, denoted as $G(g)$, is a 
function of the scaled variable 
$g\equiv g_n= (\Delta y_n^{(\alpha)}-\big<\Delta y_n^{(\alpha)}\big>)/\sigma_{\alpha}$,
where $\sigma_{\alpha}$ is the standard deviation of $\Delta y_n^{(\alpha)}$. If $\eta$
is normally distributed, $G(\eta)=G_0(\eta)=(2\pi)^{-1/2}\exp(-\eta^2/2)$, so is also $G(g)$.
This does not hold in general for distributions $G(\eta)\ne G_0(\eta)$ because the coefficients 
$C_i^{(\alpha)}$ in Eq.~(\ref{eq:deltaynalpha}), weighting the $\eta_{n-i}$'s 
differently, can yield effectively non-indentically distributed random numbers and the central 
limit theorem does not hold. To see this in a simple case, consider that $C_i^{(\alpha)}=\exp(-i/\alpha)$.
Now, if $\alpha\to 0$ then $G(g)=G(\eta)$. Also for finite $\alpha\ll1$, the distribution
$G(g)$ will not be Gaussian. In the present case, we illustrate this behavior by assuming
that $G(\eta)=\exp(-|\eta|)$. The corresponding PDF's for $g$ are shown in Fig.~\ref{fig:pdfwalks}
in the case $\alpha=-0.2$ for series of different length $N$, where $0\le n\le N$. As one 
can see from the fi\-gu\-re, $G(g)$ remains exponentially distributed attaining the form 
$G(g)\simeq A \exp(-|g|/A)$, with $A\simeq 0.7$.

%%%%%%%%%%%%%%%%%%%%%%%%%%%%%%%%%%%%% FIGURE 4 %%%%%%%%%%%%%%%%%%%%%%%%%%%%%%%%%%%%%%%%
%%%%%%%%%%%%%%%%%%%%%%%%%%%%%%%%%%%%%%%%%%%%%%%%%%%%%%%%%%%%%%%%%%%%%%%%%%%%%%%%%%%%%%%%

%----------------------------------------------------------------------------------
\subsection{Auto-correlations}

Since for a standard RW the profile behaves, in a statistical sense, as $y_n\sim n^{1/2}$, 
actually meaning that $\big<(y_n-y_{n+m})^2\big>\sim|m|$, the fractional random walk obeys
\begin{equation}
y_n^{(\alpha)} \simeq n^{1/2-\alpha},
\label{eq:fractrwH}
\end{equation}
corresponding to $\big<(y_n^{(\alpha)}-y_{n+m}^{(\alpha)})^2\big>\sim|m|^{1-2\alpha}$.
Identifying in Eq.~(\ref{eq:fractrwH}) the power of $n$ with the Hurst exponent $H$ 
\cite{Hurst:1951} we find, 
\begin{equation}
H={1\over 2}-\alpha, \qquad {\rm and} \qquad \alpha = {1\over 2} - H. 
\label{eq:Hurst}
\end{equation}
Here, we are interested in cases where $0<H<1$, yielding for $\alpha$ the range of variation 
$-1/2<\alpha<1/2$. The case $H=1/2$ corresponds to standard (uncorrelated) behavior, values 
$H>1/2$ indicate persistence or long-time autocorrelations, while values $H<1/2$ yield 
anti-persistence or negative autocorrelations.

The Hurst exponent also determines the behavior of the auto-correlation function of the 
increments $\Delta y_n^{(\alpha)}$, $\mathcal{C}_\alpha(m)$, which is given by 
$\mathcal{C}_\alpha(m)=\big< g_n g_{n+m}\big>$. If long-range memory is indeed present, 
the autocorrelation function is expected to obey the scaling behavior
$\mathcal{C}_\alpha(m)\sim |m|^{-2(1-H)}$ (see e.g.\ \cite{Peng:1994,PRL1998}), yielding
\begin{equation}
\mathcal{C}_\alpha(m)\simeq -{\rm sign}(\alpha)~|m|^{-(1+2\alpha)}, \qquad |m|\gg1.
\label{eq:autocorrel}
\end{equation}

To detect the presence of long-range memory for a fractional random walk $y_n^{(\alpha)}$, 
we apply the method known in literature as the fluctuation analysis (FA) \cite{PRL1998} based 
on Haar wavelets (HW) \cite{Kantelhardt:1995}, consisting in studying the scaling behavior 
of $y_n^{(\alpha)}$ on the time scale $t$. 
To do this, the total number of points in the series, $N$, is divided into consecutive 
non-overlaping segments of length $\ell\ge1$, corresponding to the time scale $t=\ell \tau$. 
Inside each segment $s$, $1\le s\le N/\ell$, the average of $y_n^{(\alpha)}$, denoted as 
$B_s(\ell)$, is evaluated according to
\begin{equation} 
B_s(\ell)=\frac{1}{\ell}\sum_{j=1}^{\ell} y^{(\alpha)}_{(s-1)\ell+j}.
\end{equation}
The FAHW approach consists in studying the fluctuations of the profile on the time scale 
$\ell=t/\tau$, defined as
\begin{equation}  
F(\ell) = \big <\big[ B_{s}(\ell)-B_{s-1}(\ell) \big]^2 \big>^{1/2},
\label{eq:FAHW}
\end{equation}
corresponding to the first-order Haar wavelet, and the average is performed over all consecutive 
boxes $s$ and $s-1$. To be noted is that Eq.~(\ref{eq:FAHW}) can be generalized to higher order 
wavelets \cite{PRL1998}.

%%%%%%%%%%%%%%%%%%%%%%%%%%%%%%%%%%%%%%%%% FIGURE 5 %%%%%%%%%%%%%%%%%%%%%%%%%%%%%%%%%%%%%%%%%%
\begin{figure}[t]
\vspace{0.2cm}
\begin{center}
\includegraphics[width=6.5cm]{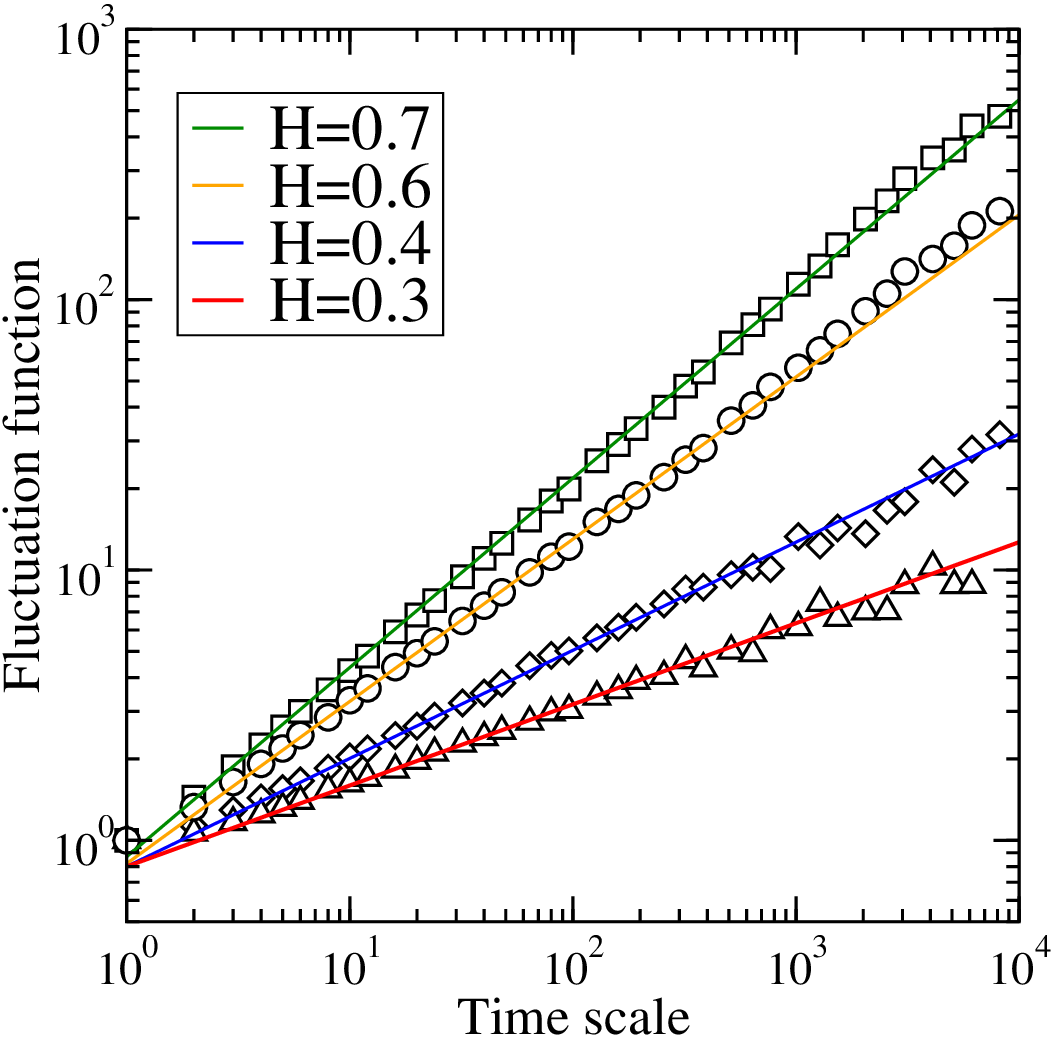}
\end{center}
\vspace{-0.5cm}
\caption{(color online) Fluctuation analysis of fractional random walks: Plotted is
         the fluctuation function $F(\ell)/F(1)$ vs time scale $\ell$ for different values of 
	 fractional order:
	 $\alpha=0.2$ (triangles), 0.1 (diamonds), -0.1 (circles) and -0.2 (squares). The
	 straight lines have slopes $H=1/2-\alpha$ given by Eq.~(\ref{eq:Hurst}). The series
         considered consist of $N=10^5$ time steps each.}
\label{fig:Hurstder}
\end{figure}
%%%%%%%%%%%%%%%%%%%%%%%%%%%%%%%%%%%%%%%%%%%%%%%%%%%%%%%%%%%%%%%%%%%%%%%%%%%%%%%%%%%%%%%%%%%%%%

The dependence of $F(\ell)$ on $\ell$ is expected to obey a scaling behavior of the form 
\begin{equation}  
F(\ell)\sim \ell^H, 
\label{eq:HurstF1}
\end{equation}
from which one can obtain the Hurst exponent $H$. Results for the fractional orders considered
in Fig.~\ref{fig:fractrandwalks} are reported in Fig.~\ref{fig:Hurstder}. The results
support the expectations of the existence of long-range memory for FRW's.

%----------------------------------------------------------------------------------
\section{Fractional random walks for financial time series}
\label{frwarch}

It is well known that daily log-returns, i.e.\ the variation of the logarithm of asset 
price $P_n$ at day $n$, denoted as $\Delta S_n=\ln P_n-\ln P_{n-1}$, is not correlated with 
its variations $\Delta S_{n'\ne n}$ and the corresponding PDF displays power-law tails, 
$G(g)\sim |g|^{-\beta}$ with $\beta\approx4$ (see e.g.\ \cite{Gopikrishnan1998,Roman2006}). 
In addition, absolute log-returns, that is $|\Delta S_n|$, appear to be long-range auto-correlated 
(see e.g.\ \cite{MantegnaStanley,FBMRomanPortoPrepr}). We aim at modeling these features using 
the FRW described above for negative values of $\alpha$.

\begin{figure}[t]
\vspace{0.1cm}
\begin{center}
\includegraphics[width=5.8cm]{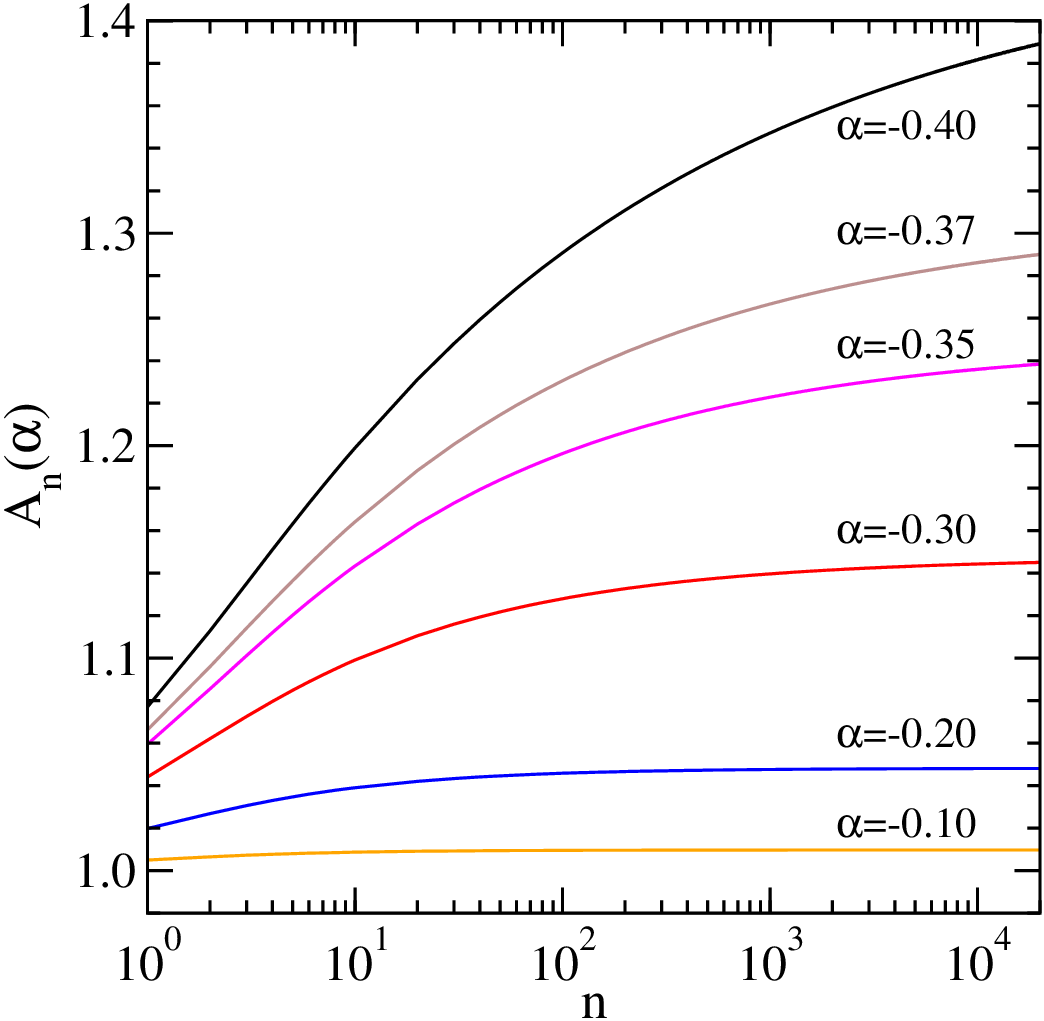}
\end{center}
\vspace{-0.5cm}
\caption{(color online) The normalization coefficients $A_n(\alpha)$ versus $n$ for
          different values of $\alpha$ (from top to bottom): $\alpha=-0.4$, $-0.37$,
	  $-0.35$, $-0.30$, $-0.20$ and $-0.10$. Notice the slow convergence of the
	  coefficient for values $\alpha\to -0.5$.}
\label{fig:ancoeff}
\end{figure}

In order to obtain a PDF for log-returns displaying power-law tails, we resort to a
simple auto-regressive conditional heteroskedastic model (ARCH) \cite{arch}. We define
the log-returns according to
\begin{equation}  
\Delta S_n = \sigma_n \eta_n, \qquad n\ge 1,
\label{eq:logretarch}
\end{equation}
where $\eta_n$ are uncorrelated random numbers drawn from a normal distribution with
zero mean and unit variance, i.e.\ $\big<\eta_n^2\big>=1$ and 
$\big<\Delta S_n\Delta S_m\big>=\big<(\Delta S)^2\big>\delta_{n,m}$. The standard 
deviation $\sigma_n$ changes in time according to the ARCH prescription,
\begin{equation}  
\sigma_n^2 = a + b~(\Delta X_{n-1}^{(\alpha)})^2,
\label{eq:sigmaArch}
\end{equation}
where $\Delta X_{n-1}^{(\alpha)}$ is proportional to the fractional integral of 
$\Delta S_{n-1}$, Eq.~(\ref{eq:deffracoper}),
\begin{equation}  
\Delta X_{n-1}^{(\alpha)} = {1\over A_{n-1}(\alpha)} D_1^{(\alpha)} \Delta S_{n-1},
                 \qquad \alpha<0,
\label{eq:deltaXArch}
\end{equation}
and the factor $A_{n-1}(\alpha)=\sqrt{\sum_{i=0}^{n-1} (C_i^{(\alpha)})^2}$ ensures the 
constancy of the 2nd moment $\big<(\Delta X_{n-1}^{(\alpha)})^2\big>$. Indeed, using 
Eq.~(\ref{eq:deffracoper}) the latter becomes
\begin{eqnarray}  
\big<(\Delta X_{n-1}^{(\alpha)})^2\big> 
&=& {1\over A^2_{n-1}(\alpha)} \sum_{i=0}^{n-1} 
    \left(C_i^{(\alpha)}\right)^2 \big<\big(\Delta S_{n-1-i}\big)^2\big>, \nonumber \\
&\equiv& \big<\big(\Delta S\big)^2\big> = \sigma_{\rm A}^2,
\label{eq:aver2deltaX}
\end{eqnarray}
where we find $\sigma_{\rm A}^2=a/(1-b)$ according to Eq.~(\ref{eq:sigmaArch}).
Plots of $A_n(\alpha)$ are shown in Fig.~\ref{fig:ancoeff}. The present combination of 
a FRW with ARCH will be denoted for brevity FRWARCH.

%%%%%%%%%%%%%%%%%%%%%%%%%%%%%%%%%%%%% FIGURE 6 %%%%%%%%%%%%%%%%%%%%%%%%%%%%%%%%%%%%%%%%
%%%%%%%%%%%%%%%%%%%%%%%%%%%%%%%%%%%%%%%%%%%%%%%%%%%%%%%%%%%%%%%%%%%%%%%%%%%%%%%%%%%%%%%% 

\begin{figure}[t]
\vspace{0.1cm}
\begin{center}
\includegraphics[width=6.cm]{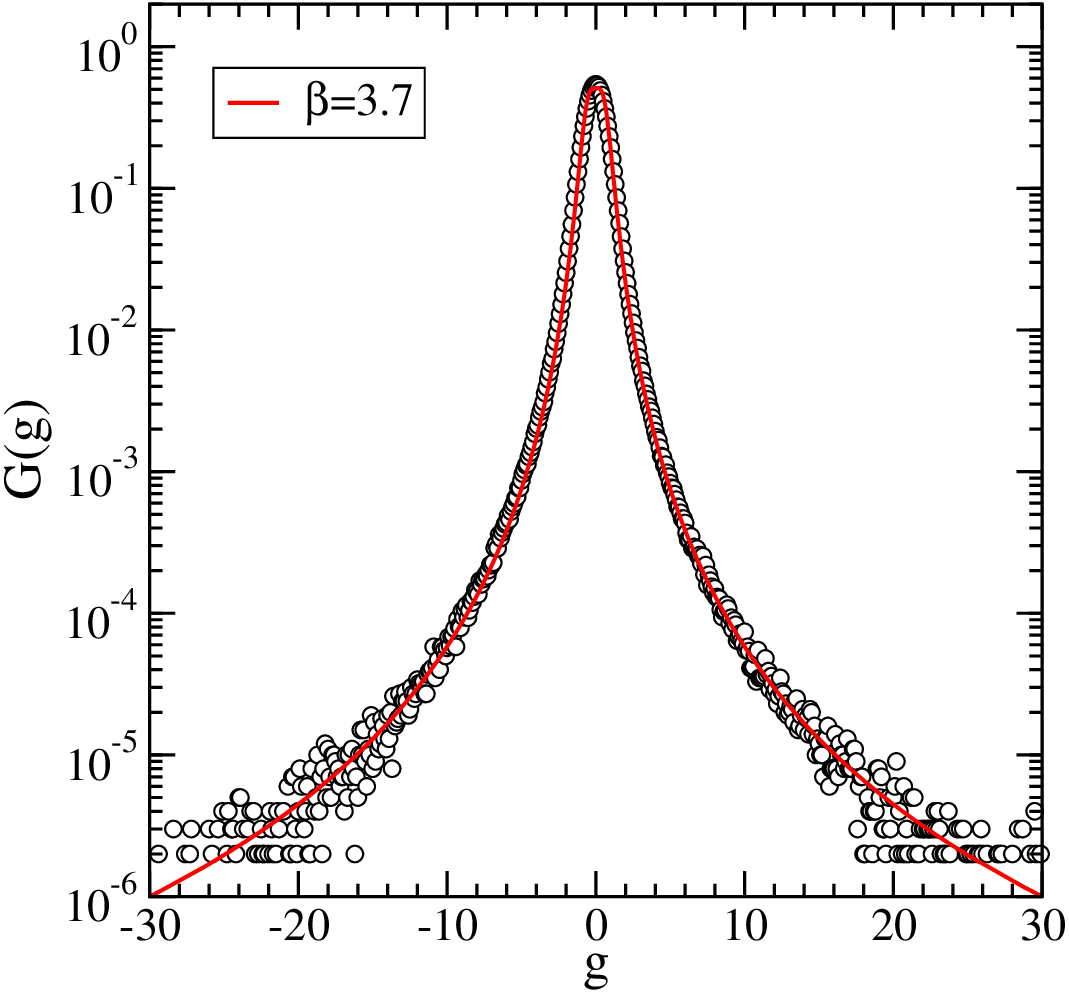}
\end{center}
\vspace{-0.5cm}
\caption{(color online) PDF, $G(g)$, for FRWARCH variations $\Delta S_n$ as a function 
          of the scaled variable $g=(\Delta S_n-\big<\Delta S_n\big>)/\sigma_{_{\rm A}}$,
          in the case $a=0.5$, $b=0.7$ and $\alpha=-0.4$. The surrogate series considered
	  consist of $N=10^{5}$ time steps each and averages over 100 configurations
	  were performed. The continuous line is a fit with the form: 
	  $f=f_0/(1+|g/g_0|^{\beta})$, with $f_0=0.51$, $g_0=0.86$ and $\beta=3.7$.
	  Here, we find $\sigma_{_{\rm A}}\simeq 1.30$, in agreement with 
	  Eq.~(\ref{eq:aver2deltaX}).}
\label{fig:pdfss} 
\end{figure}

In the following, we discuss numerical results to illustrate the implementation of
FRWARCH. We consider the fractional order $\alpha=-0.4$, together with the ARCH parameters
$a=0.5$ and $b=0.7$. To improve the accuracy of the numerically obtained PDF of $\Delta S_n$, 
we have consi\-de\-red walks of up to $N=10^5$ time steps and averaged over 100 
configurations. The results for the PDF, $G(g)$, are plotted in Fig.~\ref{fig:pdfss} as a 
function of the scaled variable $g$. A simple fit has been conducted to the numerical 
results suggesting a broad distribution function with power-law tails, $G(g)\sim |g|^{-\beta}$, 
for $|g|\gg1$, with $\beta\simeq3.7$, the latter consistent with the value expected 
analytically \cite{RomanPorto2001,DosePortoRoman2003,gammabeta}.

%%%%%%%%%%%%%%%%%%%%%%%%%%%%%%%%%%%%% FIGURE 7 %%%%%%%%%%%%%%%%%%%%%%%%%%%%%%%%%%%%%%%%
%%%%%%%%%%%%%%%%%%%%%%%%%%%%%%%%%%%%%%%%%%%%%%%%%%%%%%%%%%%%%%%%%%%%%%%%%%%%%%%%%%%%%%%%

\begin{figure}[t]
\vspace{0.1cm}
\begin{center}
\includegraphics[width=6.cm]{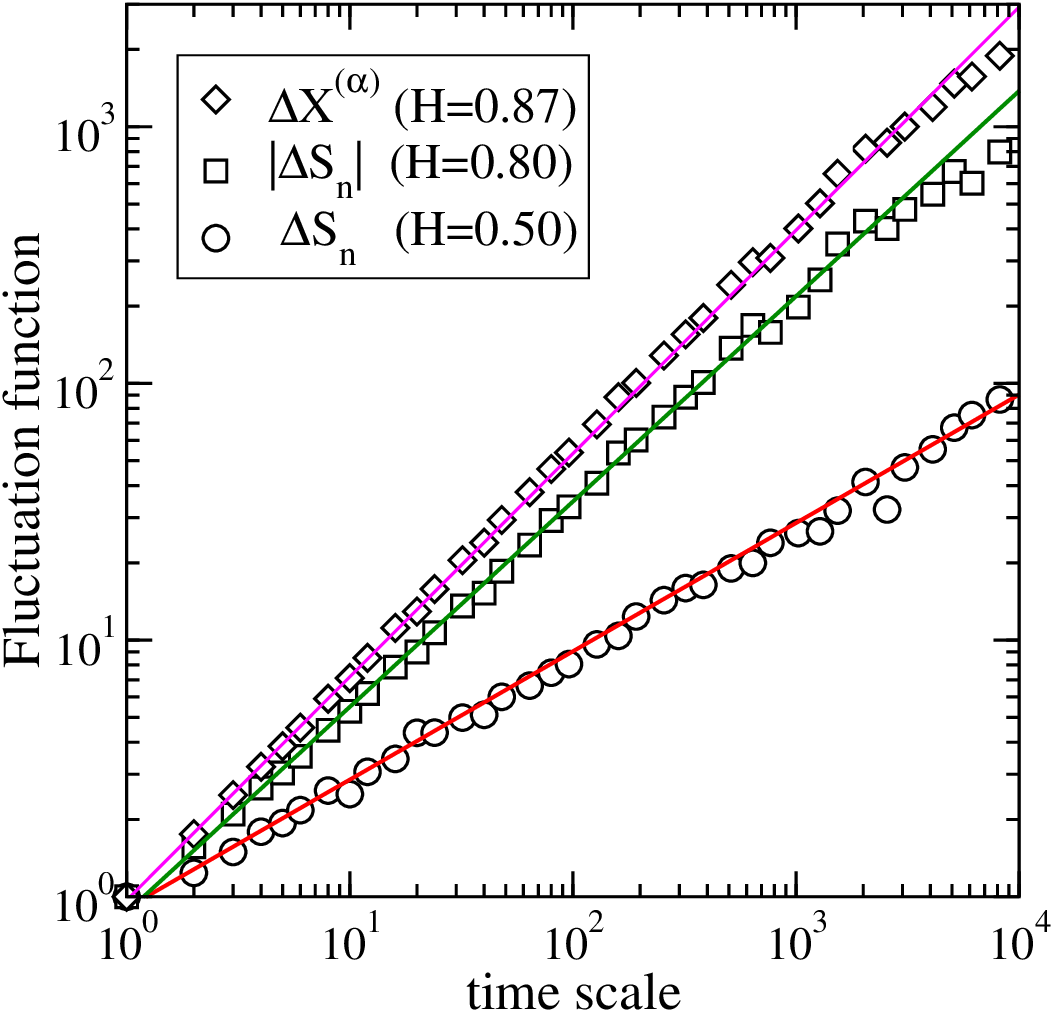}
\end{center}
\vspace{-0.5cm}
\caption{(color online) Fluctuation analysis for FRWARCH time series in the case $\alpha=-0.4$
         (see Fig.~\ref{fig:pdfss}). Plotted is the fluctuation function $F(\ell)/F(1)$ vs time
	 scale $\ell$ for: 
	 Log-returns (circles), absolute log-returns (squares) and fractional random process, 
	 Eq.~(\ref{eq:deltaXArch}), (diamonds). The straight lines have slopes $H$ as 
	 indicated in the inset. The fluctuation analysis (FAHW) was done on a single time 
	 series of length $N=10^5$.}
\label{fig:Hurstss}
\end{figure}

The fluctuation analysis of FRWARCH time series is reported in Fig.~\ref{fig:Hurstss},
where we plot the function $F(\ell)$ as a function of time scale $\ell$ for $\Delta S_n$ and
its absolute value $|\Delta S_n|$, together with the behavior of the fractional process
$\Delta X_{n}^{(\alpha)}$, Eq.~(\ref{eq:deltaXArch}). As expected, log-returns $\Delta S_n$ 
display uncorrelated behavior ($H_{\rm \Delta S}\simeq 0.5$), while the corresponding absolute 
returns show persistence on long time scales with an effective exponent 
$H_{\rm |\Delta S|}\simeq 0.8$. 

%%%%%%%%%%%%%%%%%%%%%%%%%%%%%%%%%%%%%%%%% FIGURE 8 %%%%%%%%%%%%%%%%%%%%%%%%%%%%%%%%%%%%%%%%%%
%%%%%%%%%%%%%%%%%%%%%%%%%%%%%%%%%%%%%%%%%%%%%%%%%%%%%%%%%%%%%%%%%%%%%%%%%%%%%%%%%%%%%%%%%%%%%%

We also include the results for $\Delta X_{n}^{(\alpha)}$ indicating that for the fractional 
process the effective Hurst exponent, $H_{\rm \Delta X}\simeq0.87$, is close to the expected 
value $H=1/2-\alpha=0.9$. We conclude that the long-time memory induced in absolute returns 
is a bit weaker than the one for the fractional process $\Delta X_{n}^{(\alpha)}$. However, 
the value of $H_{\rm |\Delta S|}\simeq 0.8$ is still consistent with the time decay of the 
autocorrelation function observed in empirical data for which the corresponding Hurst exponent 
typically varies within the range $0.7\lesssim H_{\rm |\Delta S|}\lesssim 0.9$ 
\cite{Roman2006,FBMRomanPortoPrepr}.

%----------------------------------------------------------------------------------
\section{FIGARCH revisited}
\label{figarch}

We briefly review the main properties of FIGARCH models \cite{Baillie1996} relevant
for our present discussion. Let us start with the GARCH model \cite{Bollerslev:1986}, 
which is considered here in its simplest form. It is defined analogously to 
Eqs.~(\ref{eq:logretarch}) and (\ref{eq:sigmaArch}), where now
\begin{equation}
\sigma_{n}^2 = a + b~(\Delta S_{n-1})^2 + c~\sigma_{n-1}^2,
\label{eq:sigma2garch}
\end{equation}
and $a,b,c$ are positive constants. The mean variance becomes
\begin{equation}
\sigma^2_{\mathrm{G}} = \frac{a}{1-(b+c)},
\label{eq:garchsigma2}
\end{equation}
which is finite provided that $b+c<1$. In order to discuss the FIGARCH model, one
resorts to the fractional differencing operator, defined as
\begin{eqnarray}
\label{eq:fractdiffoper}
(1-L)^\theta &=& 1- \theta L - \frac{1}{2!} \theta (1-\theta) L^2           \nonumber\\
       & & ~~~~~~~~~ - \frac{1}{3!} \theta (1-\theta) (2-\theta) L^3, \dots \nonumber\\
             &=& 1 - \sum_{i=1}^\infty {\rm C}_i(\theta) L^i,	    
\end{eqnarray}
where $0<\theta<1$, ${\rm C}_i(\theta)>0$ and $L$ is the lag operator, defined according to
$L^i x^2_{n}=x^2_{n-i}$. Analogously to Eqs.~(\ref{eq:binomcoefffact}), (\ref{eq:expancoeff}) 
and (\ref{eq:asympcoeff}), the coefficients ${\rm C}_i(\theta)$ above behave for large $i$
according to
\begin{equation}
{\rm C}_i(\theta)\sim {\theta \over \Gamma(1-\theta)} i^{-(1+\theta)}, \qquad i\gg1.
\label{eq:cjasymp}
\end{equation}
Note that ${\rm C}_i(\theta) = -C_i^{(\alpha)}$, for $\alpha>0$, as can be seen from
Eq.~(\ref{eq:expancoeff}). Finally, taking $L=1$ in Eq.~(\ref{eq:fractdiffoper}), 
one finds the general sum-rule, $R(\theta)=\sum_{i=1}^\infty {\rm C}_i(\theta)=1$, 
valid for $0<\theta<1$.

The standard way of introducing FIGARCH is to write Eq.~(\ref{eq:sigma2garch}) using
the lag operator $L$ in the form
\begin{equation}
(1-bL-cL) (\Delta S_{n})^2 = a + (1-c L) [(\Delta S_{n})^2 - \sigma_{n}^2],
\label{eq:lagsigma2garch}
\end{equation}
and inserting the differencing operator $(1-L)^\theta$ in the left-hand side of the 
above relation, yielding,
$$(1-bL-cL) (1-L)^\theta (\Delta S_{n})^2 = a + (1-c L) [(\Delta S_{n})^2 - \sigma_{n}^2].$$
This is a generalization of the $\theta=1$ integrated GARCH (or IGARCH) model 
\cite{Baillie1996} to the case of a fractional exponent $0<\theta<1$.

Expanding the operator $(1-L)^\theta$ according to Eq.~(\ref{eq:fractdiffoper}), we find
\begin{eqnarray}
\label{eq:sigma2figarch}
\hspace{-1cm}
\sigma_{n}^2 &=& a + b~(\Delta  S_{n-1})^2 + c~\sigma_{n-1}^2            \\
             &+& \sum_{i=1}^{\infty} {\rm C}_i(\theta) \left[ 
                (\Delta S_{n-i})^2-(b+c)(\Delta S_{n-1-i})^2  \right],  \nonumber 
\end{eqnarray}
which, using the relation ${\rm C}_{i-1}(\theta)/{\rm C}_{i}(\theta)=i/(i-1-\theta)$, 
becomes
\begin{eqnarray}
\label{eq:sigma2figarchnew}
\hspace{-1cm}
\sigma_{n}^2 &=& a + (b+\theta)~(\Delta S_{n-1})^2 + c~\sigma_{n-1}^2      \nonumber \\
&+& \sum_{i=2}^{\infty} {\rm C}_i(\theta)~f_i(\theta,b,c)~(\Delta S_{n-i})^2,    
\end{eqnarray}
where
\begin{equation}
f_i(\theta,b,c) = {(i-1)(1-b-c)-(\theta+b+c)\over i-1-\theta}.
\label{eq:condfi}
\end{equation}
Stability of the process is achieved when the coefficients $f_i(\theta,b,c)$,
Eq.~(\ref{eq:condfi}), are positive yielding the minimal condition
\begin{equation}
\theta+2(b+c) < 1.
\label{eq:condfifarch}
\end{equation}

 From Eq.~(\ref{eq:sigma2figarch}), we can write the mean variance for FIGARCH as
\begin{equation}
\sigma^2_{\mathrm{F}} = \frac{a}{[1-(b+c)]~[1-R(\theta)]},
\label{eq:figarchsigma2}
\end{equation}
which diverges for all $0<\theta\le 1$, because $R(\theta)=1$. In the case $0<\theta<1/2$, 
the autocorrelation function of $\sigma_n^2$ for a FIGARCH process decays as the power-law
\cite{Baillie1996}
\begin{equation}
\mathcal{C}(\tau)\sim |\tau|^{-(1-2\theta)}, \qquad |\tau|\gg 1.
\end{equation}
Writing the latter as $\mathcal{C}(\tau)\sim |\tau|^{-2(1-H)}$ (see above 
Eq.~(\ref{eq:autocorrel})), we find the relation $\theta=H-1/2$.

In practical calculations, the sum in Eq.~(\ref{eq:sigma2figarchnew}) has a finite number of
terms, with an upper memory cut-off that we denote as $M_0$. For finite $M_0$, the sum 
$R(\theta)<1$ and as a result the mean variance of the process is finite. In order to
improve the convergence of the finite $M_0$ case to FIGARCH, one can use renormalized
coefficients \cite{Baillie1996,Zumbach2004},
\begin{equation}
\tilde{\rm C}_i(\theta)={{\rm C}_i(\theta)\over \sum_{i=1}^{M_0}{\rm C}_i(\theta)},
\label{eq:ctilde}
\end{equation}
so that $\tilde{R}(\theta)=\sum_{i=1}^{M_0} \tilde{\rm C}_i(\theta)=1$. The final expression
for $\sigma_n^2$ then reads,
\begin{eqnarray}
\label{eq:sigma2figarchpract}
%\hspace{-1cm}
\sigma_{n}^2 &=& a + (b+\tilde{\theta})~(\Delta S_{n-1})^2 + c~\sigma_{n-1}^2           \\
&+& \sum_{i=2}^{M_0} \tilde{\rm C}_i(\theta)~f_i(\theta,b,c)~(\Delta S_{n-i})^2,    \nonumber
\end{eqnarray}
where $\tilde{\theta}=\theta/\sum_{i=1}^{M_0}{\rm C}_i(\theta)$.

%%%%%%%%%%%%%%%%%%%%%%%%%%%%%%%%%%%%% FIGURE 9 %%%%%%%%%%%%%%%%%%%%%%%%%%%%%%%%%%%%%%%%
\begin{figure}[t]
\vspace{0.cm}
\begin{center}
\includegraphics[width=6.5cm]{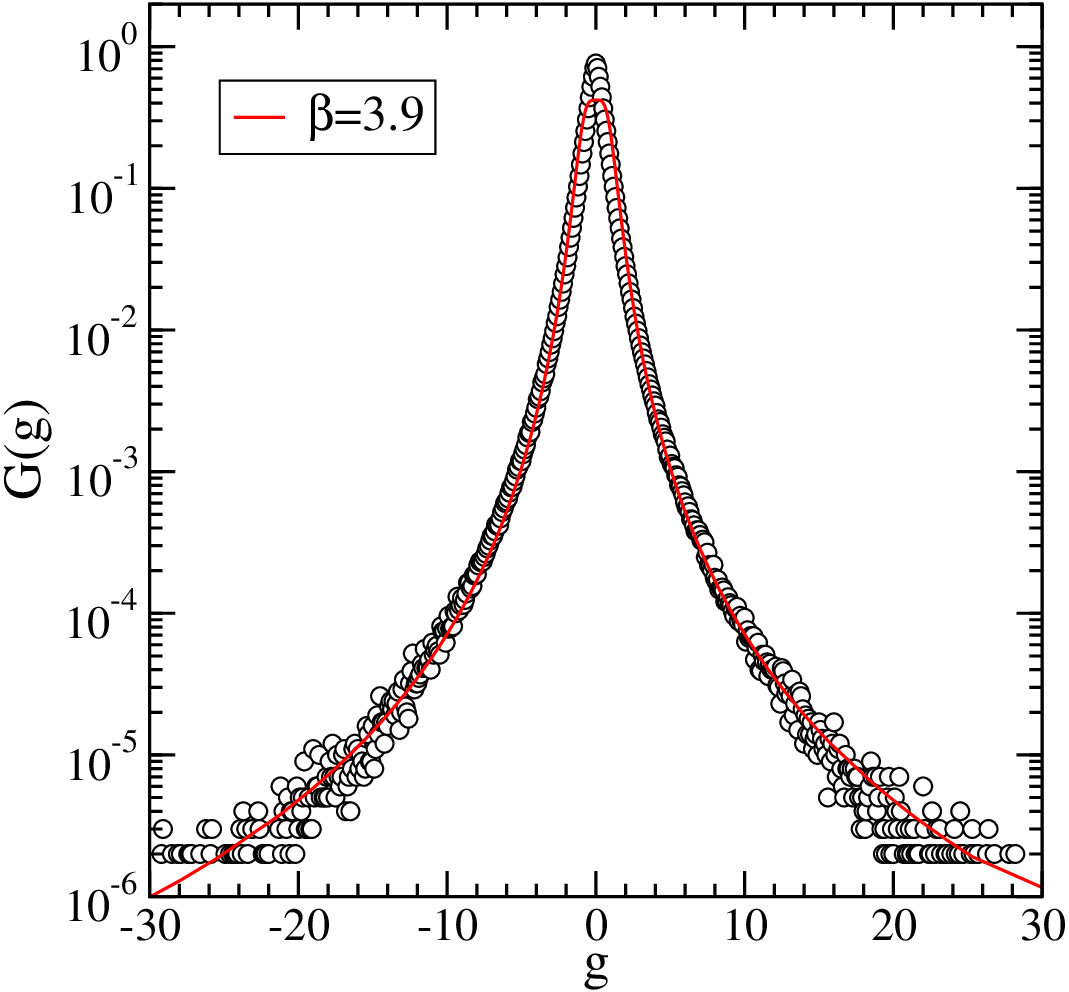}
\end{center}
\vspace{-0.5cm}
\caption{(color online) PDF, $G(g)$, for FIGARCH variations $\Delta S_n$ as a function 
          of the scaled variable $g=(\Delta S_n-\big<\Delta S_n\big>)/\sigma_{_{\rm F}}$,
          in the case $a=0.01$, $b=0.33$ and $\theta=0.3$. The surrogate series considered
	  consist of $N=10^5$ time steps each, with a memory cut-off of $M_0=10^5$ time steps. 
	  (Note that each time series is actually $2 N$ long, so that each of the $N$ time 
	  steps considered in the average process is a function of its previous $N$ steps.)
	  Averages over 100 configurations were performed. The continuous line has the form: 
	  $f=f_0/(1+|g/g_0|^{\beta})$, with $f_0=0.42$, $g_0=1.08$ and $\beta=3.9$. Here, 
	  we find $\sigma_{_{\rm F}}\simeq 1.20$ and $\tilde{\theta}=0.307$.} 
\label{fig:pdfFIG}  
\end{figure}
%%%%%%%%%%%%%%%%%%%%%%%%%%%%%%%%%%%%%%%%%%%%%%%%%%%%%%%%%%%%%%%%%%%%%%%%%%%%%%%%%%%%%%%%

In order to make contact with the FRW results discussed in Sect.~\ref{frwarch}, we consider
the simpler case $c=0$ in Eq.~(\ref{eq:sigma2figarchnew}), and take $a=0.01$ (in order to
obtain a mean standard deviation similar to FRWARCH), $b=0.33$ (in order to get a power-law
PDF as in Fig.~\ref{fig:pdfss}) and $\theta=0.3$. The latter is chosen to yield the same value 
of $H=0.8$ characterizing the autocorrelation of absolute returns obtained from FRWARCH in 
the case $\alpha=-0.4$. Note also that the chosen value of $b$ obeys the boundary condition 
Eq.~(\ref{eq:condfifarch}).

\begin{figure}[t]
\vspace{0.1cm}
\begin{center}
\includegraphics[width=6.5cm]{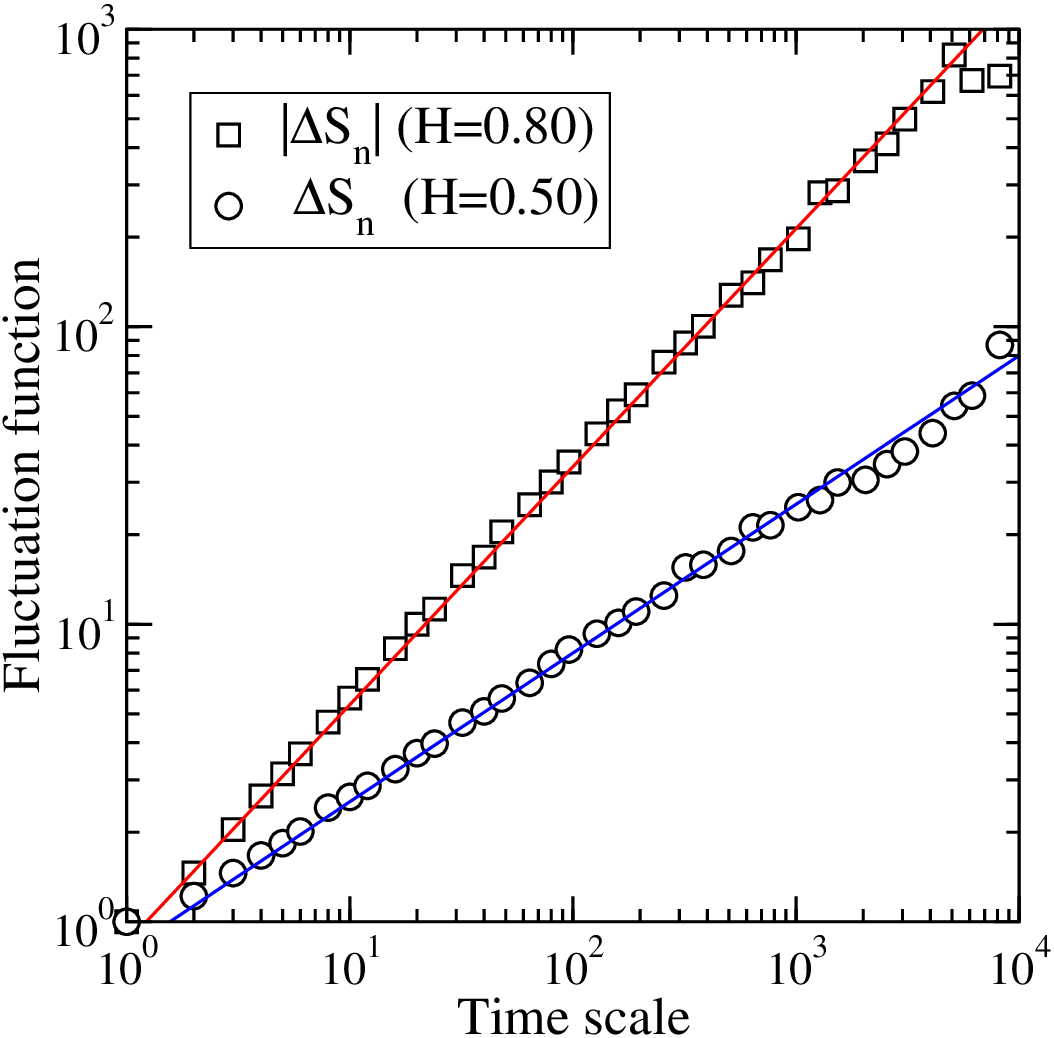}
\end{center}
\vspace{-0.5cm}
\caption{(color online) Fluctuation analysis for FIGARCH time series in the case $\theta=0.3$.
         Plotted is the fluctuation function $F(\ell)/F(1)$ vs time scale $\ell$ for: 
	 Log-returns (circles) and absolute log-returns (squares). The straight lines have slopes 
	 $H$ as indicated in the inset. The fluctuation analysis (FAHW) was done on a single 
	 time series of length $N=10^5$ with $M_0=10^5$ steps.}
\label{fig:HurstFIG}
\end{figure}

Results for the PDF of $\Delta S_n$, obtained in the case $N=M_0=10^5$, are shown in scaled 
form in Fig.~\ref{fig:pdfFIG}. Also here, $G(g)\sim |g|^{-\beta}$, for $|g|\gg1$, with 
$\beta\simeq3.9$, similar to FRWARCH in Fig.~\ref{fig:pdfss}. Finally, results of the 
fluctuation analysis are displayed in Fig.~\ref{fig:HurstFIG}, supporting the expectation 
that for log-returns no correlations are present, while for absolute log-returns the effective 
Hurst exponent is consistent with $H=1/2+\theta=0.8$.

%%%%%%%%%%%%%%%%%%%%%%%%%%%%%%%%%%%%%%%%% FIGURE 10 %%%%%%%%%%%%%%%%%%%%%%%%%%%%%%%%%%%%%%%%%%
%%%%%%%%%%%%%%%%%%%%%%%%%%%%%%%%%%%%%%%%%%%%%%%%%%%%%%%%%%%%%%%%%%%%%%%%%%%%%%%%%%%%%%%%%%%%%%

%----------------------------------------------------------------------------------
\section{FIARCH revisited}
\label{fiarch}

\begin{figure}[t]
\vspace{0.1cm}
\begin{center}
\includegraphics[width=6.5cm]{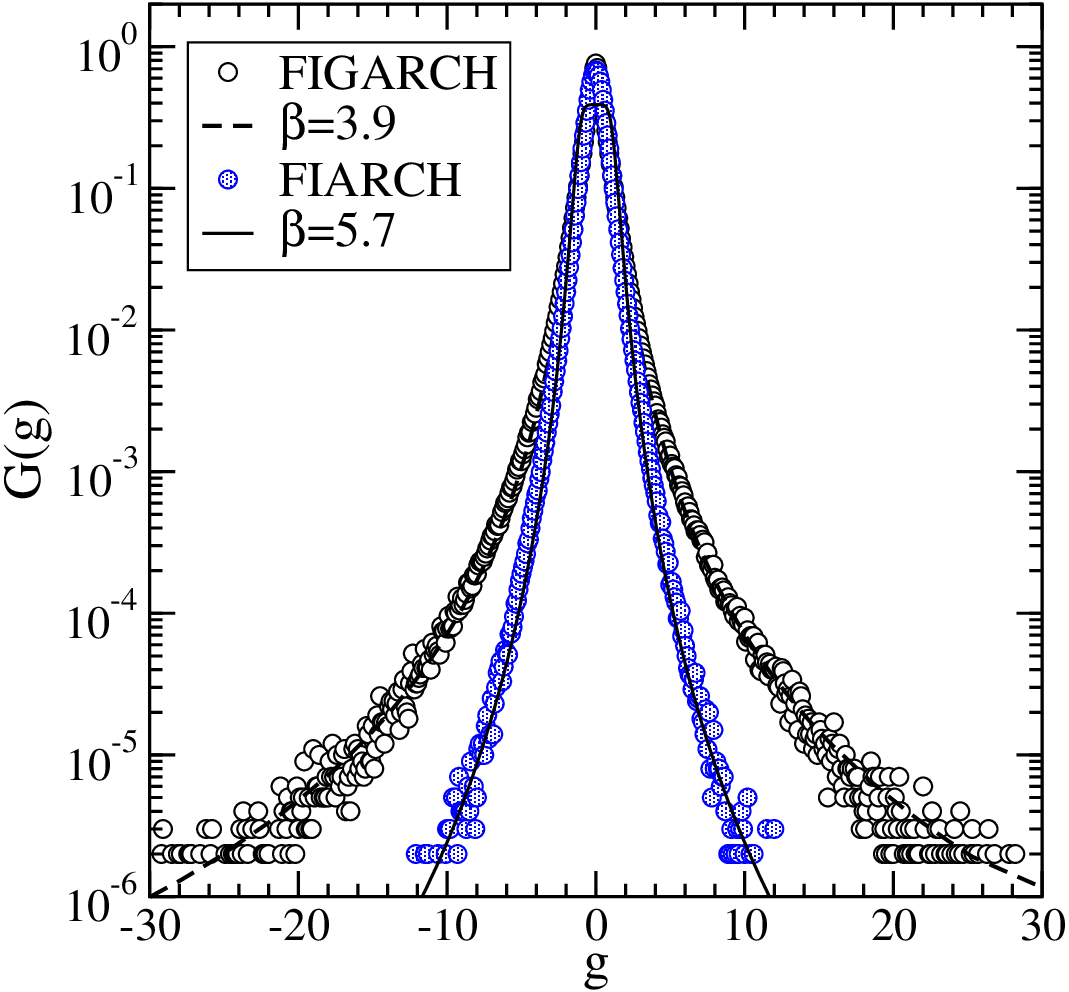}
\end{center}
\vspace{-0.5cm}
\caption{(color online) PDF, $G(g)$, for FIARCH variations $\Delta S_n$ as a function 
          of the scaled variable $g=(\Delta S_n-\big<\Delta S_n\big>)/\sigma_{_{\rm F}}$,
          in the case $\sigma_0=1.2$ and $\theta=0.3$ (full circles). The FIARCH series 
	  consisted of $N=10^5$ time steps each, with a memory cut-off of $M_0=10^5$ time 
	  steps. Averages over 100 configurations were performed. The continuous line has 
	  the form: $f=f_0/(1+|g/g_0|^{\beta})$, with $f_0=0.39$, $g_0=1.22$ and $\beta=5.7$. 
	  Here, we find $\big<\sigma_{n}^2\big>^{1/2}\simeq 1.30$ and $\tilde{\theta}=0.307$.
	  For comparison, results from FIGARCH (open circles, taken from Fig.~\ref{fig:pdfFIG}) 
	  have been included.} 
\label{fig:pdfFIARCH}  
\end{figure}

There are other models currently used in literature regarding long-time memory of absolute
log-returns and variable volatility. We briefly comment on them in order to better assess 
the impact of FRWARCH suggested here. Let us consider fractionally integrated ARCH process 
\cite{GrangerDing:1996}, variants of which have been studied recently \cite{Podobnik2006,Podobnik2007}. 
The model is based on the equations
\begin{eqnarray}  
\label{eq:fiarchrev1}
\Delta S_n &=& \sigma_n \eta_n, \qquad n\ge 1,                               \\
\sigma_n   &=& \sigma_0 \sum_{i=1}^{M_0} \tilde{\rm C}_i(\theta) 
                 {|\Delta S_{n-i}|\over \big<|\Delta S_{n}|\big>},
\label{eq:fiarchrev}
\end{eqnarray}
where $\eta_n$ are independent normally distributed random numbers ($\big<\eta_n^2\big>=1$)
and $0<\theta<1/2$ as for FIGARCH. According to the sum rule 
$\sum_{i=1}^{M_0} \tilde{\rm C}_i(\theta)=1$, see Eq.~(\ref{eq:ctilde}), Eq.~(\ref{eq:fiarchrev}) 
yields $\big< \sigma_n\big>=\sigma_0$. Thus, log-returns $\Delta S_n$ are uncorrelated to each other
while absolute returns, $|\Delta S_n|$, display long-time memory. As one can see from
Eq.~(\ref{eq:fiarchrev}), the only free parameters are $\theta$ and $\sigma_0$ \cite{Mzero}, in 
contrast to the three parameters for FRWARCH. This has consequences on the shape of the PDF as we 
can see from the numerical results reported in Fig.~\ref{fig:pdfFIARCH}.

%%%%%%%%%%%%%%%%%%%%%%%%%%%%%%%%%%%%% FIGURE 11 %%%%%%%%%%%%%%%%%%%%%%%%%%%%%%%%%%%%%%%%
%%%%%%%%%%%%%%%%%%%%%%%%%%%%%%%%%%%%%%%%%%%%%%%%%%%%%%%%%%%%%%%%%%%%%%%%%%%%%%%%%%%%%%%%

The resulting PDF for FIARCH is indeed consistent with a power-law distribution at the tails, 
but the power-law exponent turns out to be large, here $\beta\approx6$, and can not be 
controlled by tuning a model parameter. Indeed, the value of $\theta=0.3$ is fixed by the condition
that $H=0.8$. Results of fluctuation analysis for FIARCH (not shown here) confirm that
in this case $H\approx0.8$. Thus, it appears that additional degrees of freedom, in terms
of model parameters, are required in Eq.~(\ref{eq:fiarchrev}) in order to obtain PDF
with varying shapes.

To do this, we suggest a slight generalization of Eq.~(\ref{eq:fiarchrev}) to the form,
\begin{equation}  
\sigma_n = \sigma_0 \sum_{i=1}^{M_0} \tilde{\rm C}_i(\theta) 
           {|\Delta S_{n-i}|\over \big<|\Delta S_{n}|\big>} +  b~|\Delta S_{n-1}|,
\label{eq:fiarchmod}
\end{equation}
where the mean standard deviation now can be obtained as, 
$\big<\sigma_n\big> = \sigma_0 + b~\big<|\Delta S_{n-1}|\big> = 
\sigma_0 + b~\big<\sigma_{n-1}\big> \big<|\eta_{n-1}|\big>$,
yielding
\begin{equation}  
\big<\sigma_n\big> = {\sigma_0 \over 1-b~\big<|\eta_n|\big>}.
\label{eq:fiarch2meansig}
\end{equation}
Note that for a normal distribution $\big<|\eta_n|\big>=\sqrt{2/\pi}\approx 0.8$.

%%%%%%%%%%%%%%%%%%%%%%%%%%%%%%%%%%%%% FIGURE 12 %%%%%%%%%%%%%%%%%%%%%%%%%%%%%%%%%%%%%%%%
\begin{figure}[t]
\vspace{0.1cm}
\begin{center}
\includegraphics[width=6.5cm]{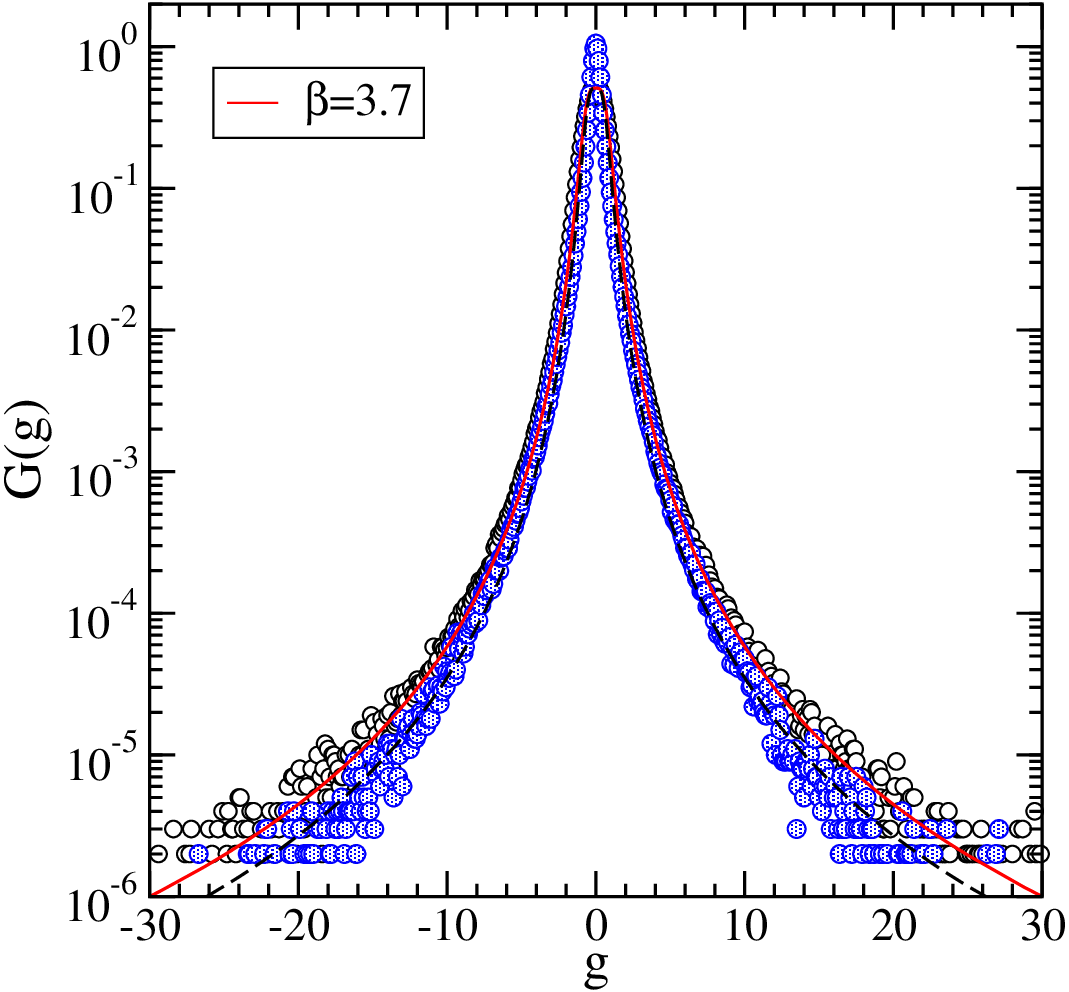}
\end{center}
\vspace{-0.5cm}
\caption{(color online) PDF, $G(g)$, for modified FIARCH variations $\Delta S_n$ as a function 
          of the scaled variable $g=(\Delta S_n-\big<\Delta S_n\big>)/\sigma_{_{\rm F}}$,
          in the case $\sigma_0=0.67$, $b=0.38$ and $\theta=0.3$ (full circles). The FIARCH series 
	  consisted of $N=10^5$ time steps each, with a memory cut-off of $M_0=10^5$ time 
	  steps. Averages over 100 configurations were performed. The dashed line has 
	  the form: $f=f_0/(1+|g/g_0|^{\beta})$, with $f_0=0.51$, $g_0=0.75$ and $\beta=3.7$. 
	  Here, we find $\big<\sigma_{n}^2\big>^{1/2}\simeq 1.30$ and $\tilde{\theta}=0.307$.
	  For comparison, results from FRWARCH (open circles and continuous line, taken 
	  from Fig.~\ref{fig:pdfss}) have been included.} 
\label{fig:pdfFIARCHmod}  
\end{figure}
%%%%%%%%%%%%%%%%%%%%%%%%%%%%%%%%%%%%%%%%%%%%%%%%%%%%%%%%%%%%%%%%%%%%%%%%%%%%%%%%%%%%%%%%

Although the additional parameter $b$ actually helps in getting different decaying
power-law exponents $\beta$ for the PDF, the model behaves similarly as FIGARCH in the
sense that there is no an a priori way to estimate $b$ and this has to be done case by
case. The numerical examples investigated (see Fig.~\ref{fig:pdfFIARCHmod}) suggest that 
appropriate set of parameters can be found but at the expense of a numerical search.

%----------------------------------------------------------------------------------
\section{Comparison to empirical data}
\label{empirical}

\begin{figure}[t]
\vspace{0.1cm}
\begin{center}
\includegraphics[width=6.5cm]{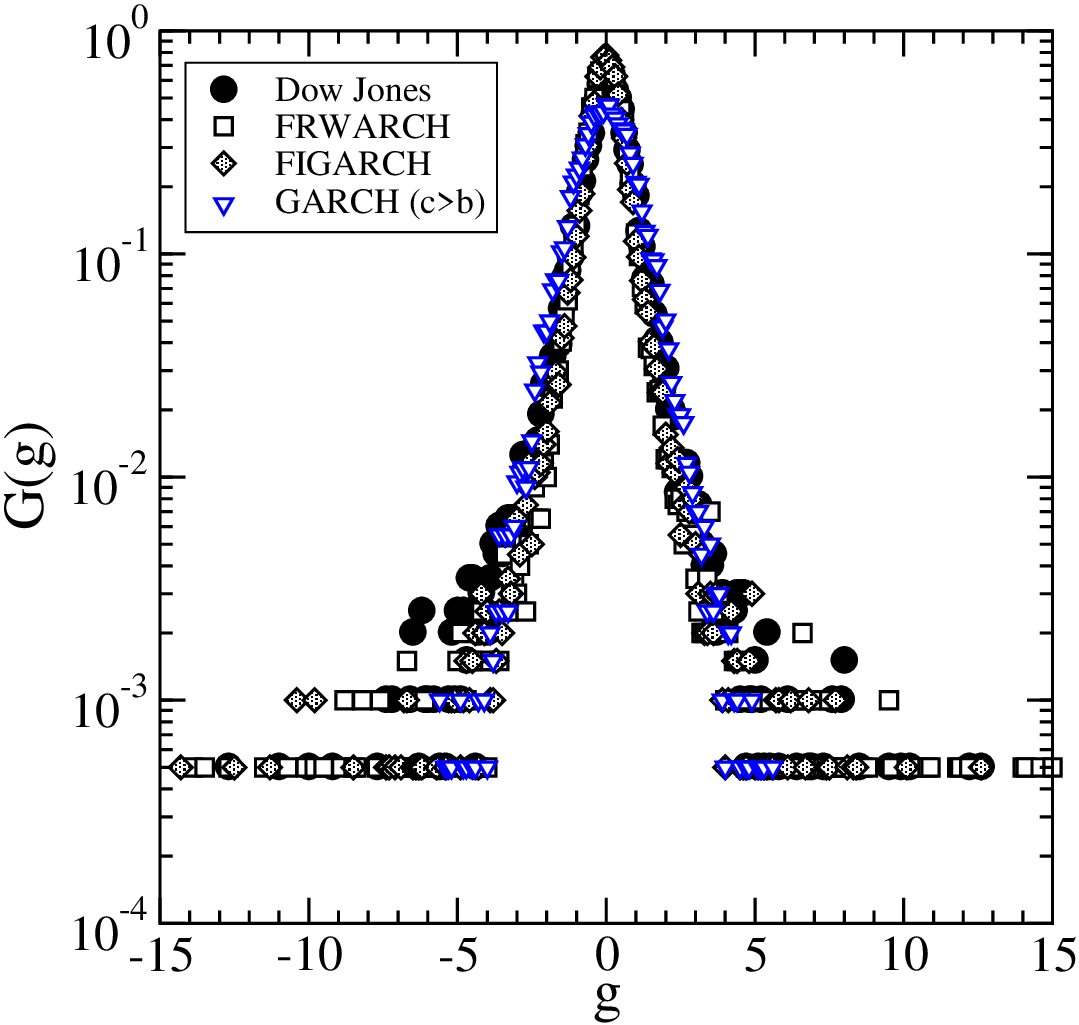}
\end{center}
\vspace{-0.5cm}
\caption{(color online) PDF, $G(g)$, for variations $\Delta S_n$ as a function of the 
          scaled variable $g=(\Delta S_n-\big<\Delta S_n\big>)/\sigma$ for: Dow Jones Index
	   (full circles), FRWARCH (open squares), FIGARCH (filled diamonds), and GARCH
	   (open triangles down). The model parameters are indicated in the text.} 
\label{fig:pdffrwfigtarch}  
\end{figure}

In this section, we discuss results for FRWARCH and FIGARCH with regards to empirical
data. For the latter, we take daily data for the Dow Jones Index, $I_{\rm DJ}(n)$, from 1st 
October 1928 till 16 May 2008 \cite{yahoo}, consisting of $19994$ values. As the working 
variable, we consider the daily changes of the logarithm of the index,
$\Delta S_n\equiv \ln I_{\rm DJ}(n) - \ln I_{\rm DJ}(n-1)$. The corresponding PDF is
shown by the full circles in Fig.~\ref{fig:pdffrwfigtarch}. To be noted is that
non-stationarity issues may play a role for the Dow Jones Index as discussed in
\cite{PlovdivRomanPorto}. Here, we disregard such corrections and assume the series
as stationary.

In the following we consider series of $N=20000$ values. Regarding FRWARCH, we use: 
$a=0.5$, $b=0.7$ and $\alpha=-0.4$. FIGARCH results are obtained for the case: 
$a=0.015$, $b=0.2$ and $\theta=0.4$ (here also $c=0$). In addition to these long-memory 
models, we consider for illustration a short-memory model such as the GARCH version 
discussed in Eq.~(\ref{eq:sigma2garch}), for the model parameters: $a=0.2$, $b=0.09$ 
and $c=0.9$. Alternatively, we will also discuss the set $b=0.9$ and $c=0.09$. 
Results for the PDF's are shown in Fig.~\ref{fig:pdffrwfigtarch}. To be noted is that
both FRWARCH and FIGARCH yield PDF's in good agreement with the empirical one. The results
from GARCH are not that satisfactory, for the chosen set of paramters. If we take instead
the alternative set ($b=0.9$ and $c=0.09$), the agreement becomes comparable to that from
FRWARCH and FIGARCH. The reason for the first choice is based on the behavior of the
fluctuation function for absolute returns, as we see below.

%%%%%%%%%%%%%%%%%%%%%%%%%%%%%%%%%%%%% FIGURE 13 %%%%%%%%%%%%%%%%%%%%%%%%%%%%%%%%%%%%%%%%
%%%%%%%%%%%%%%%%%%%%%%%%%%%%%%%%%%%%%%%%%%%%%%%%%%%%%%%%%%%%%%%%%%%%%%%%%%%%%%%%%%%%%%%%

\begin{figure}[t]
\vspace{0.1cm}
\begin{center}
\includegraphics[width=6.5cm]{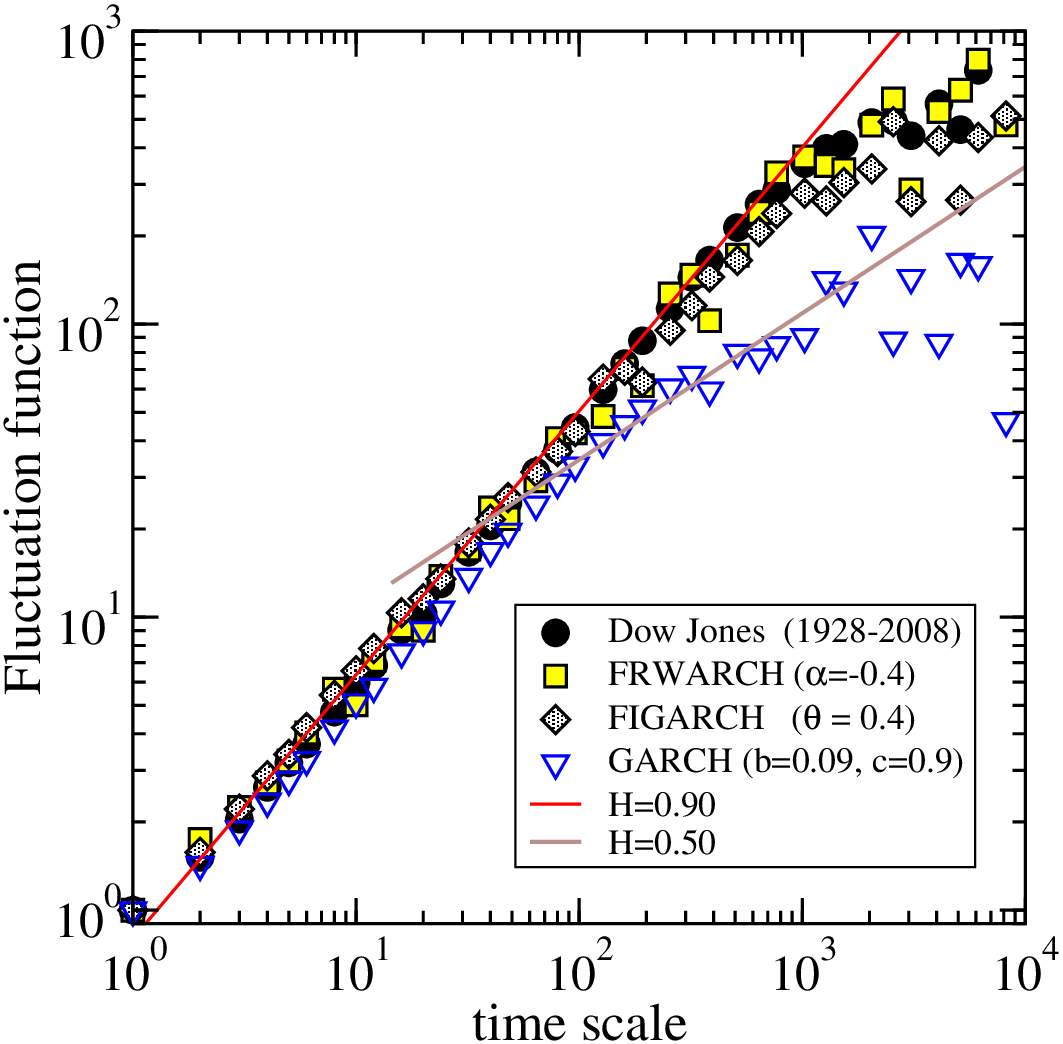}
\end{center}
\vspace{-0.5cm}
\caption{(color online) Fluctuation analysis versus time scale [days] of absolute returns, 
         for: Dow Jones Index (full circles), FRWARCH (open squares), FIGARCH (filled diamonds), 
         and GARCH (open triangles down). The straight lines have slopes $H=0.9$ and $H=0.5$,
	   and are displayed for illustration.} 
\label{fig:hurstfrwfigtarch}  
\end{figure}

The fluctuation function of absolute log-returns for the Dow Jones Index is displayed by 
the full circles in Fig.~\ref{fig:hurstfrwfigtarch}. We find power-law behavior over about 
three decades with a Hurst exponent $H\simeq 0.9$, indicating a strong persistence in the 
autocorrelation function of absolute log-returns. Similar behavior is displayed by 
$|\Delta S_n|$ for FRWARCH (open squares), where the latter almost overlap with the
empirical ones. Also FIGARCH yields results in good agreement with the Dow Jones values,
although the agreement is not as good as for FRWARCH. In contrast, GARCH yields good
results only for time scales below about 100 days. On larger times, the GARCH fluctuation 
function displays uncorrelated behavior with exponent $H=1/2$, indicating the existence of
only short-range autocorrelations in the time series, as expected. If we use the second
set of values (i.e. $b=0.9$ and $c=0.09$) for GARCH, the PDF improves considerably, but
the fluctuation function crosses over the uncorrelated regime ($H=1/2$) already at about
10 days, yielding a poor fluctuation behavior. These results suggest that a 'long-range'
memory is a necessary ingredient in a surrogate model, as the one described by FRWARCH 
or FIGARCH.

%%%%%%%%%%%%%%%%%%%%%%%%%%%%%%%%%%%%% FIGURE 14 %%%%%%%%%%%%%%%%%%%%%%%%%%%%%%%%%%%%%%%%
%%%%%%%%%%%%%%%%%%%%%%%%%%%%%%%%%%%%%%%%%%%%%%%%%%%%%%%%%%%%%%%%%%%%%%%%%%%%%%%%%%%%%%%%

%----------------------------------------------------------------------------------
\section{Conclusions}
\label{conclu}

Fractional derivatives represent a conceptually simple scheme that allow us to build,
from a standard random walk, stochastic processes with long-range autocorrelations.
The long-memory built into the walks is controlled by the order of the fractional
operator. Positive orders correspond to fractional derivatives and negative ones to 
fractional integrations. The former lead to fractional random walks (FRW) with negative 
(or anti-persistent)
autocorrelations and the latter to positive (or persistent)
autocorrelations. Long-time autocorrelations decay as a power-law for long time lags,
the exponent of which depends on the Hurst exponent associated to the walks, the
latter being a function of the fractional operator order. Examples have been studied
to illustrate the use of fluctuation analysis based on Haar wavelets to determine
the corresponding Hurst exponents. The results indicate that a constant Hurst exponent
is consistent with a wide range of time scales, suggesting that FRW are essentially 
stationary for most practical purposes. A FRW model has been discussed for describing
the behavior of both log-returns and their absolute values observed in empirical data
of financial assets, which is based on a simple autoregressive (ARCH) scheme and denoted as 
FRWARCH. Absolute log-returns, as well as volatility, display strong autocorrelations, 
and the proposed FRW model seems to capture the essential features. Statistical properties 
of the present model have been compared with the predictions of a FIGARCH and FIARCH 
processes, to illustrate the difficulties that are found in practical calculations. We 
may conclude that FRWARCH turns out to be as accurate as FIGARCH regarding long-time memory 
features and it appears to be more stable than the latter with regard to distribution 
functions of log-returns. We therefore suggest that FRWARCH is suitable for simulating 
empirically observed slowly-decaying absolute log-returns autocorrelations, competing
with the presently available models in the financial literature, as also demonstrated 
by a direct confrontation with daily close data from the Dow Jones Index.

%----------------------------------------------------------------------------------
\newpage
%----------------------------------------------------------------------------------

%%%%%%%%%%%%%%%%%%%%%%%%%%%%%%%%%%%%%%%%%%%%%%%%%%%%%%%%%%%%%%%%%%%%%%%%%%%%%%%%%%%%%%%%%%%%%%%%
%%%%%%%%%%%%%%%%%%%%%%%%%%%%%%%%%%%%%%%%%%%%%%%%%%%%%%%%%%%%%%%%%%%%%%%%%%%%%%%%%%%%%%%%%%%%%%%%

\begin{thebibliography}{99}
\bibitem{Mzero} The parameter $M_0$ is not free in the sense that it should be taken as large
                as possible in order that the results become independent of its value.
\end{thebibliography}
\end{document}